\documentclass[iop]{emulateapj}

\usepackage{color}
\usepackage{natbib}

\slugcomment{Accepted by ApJ}

\shorttitle{Simulations of the origin and fate of the Galactic Center cloud G2}
\shortauthors{Schartmann et al.}

\begin{document}


\title{Simulations of the origin and fate of the Galactic Center cloud G2}


\author{M. Schartmann\altaffilmark{1}\altaffilmark{2}, A. Burkert\altaffilmark{1}\altaffilmark{2}\altaffilmark{3}, C. Alig\altaffilmark{1}\altaffilmark{2},
S. Gillessen\altaffilmark{2}, R. Genzel\altaffilmark{2}, F. Eisenhauer\altaffilmark{2} and T.K. Fritz\altaffilmark{2}}

\email{schartmann@mpe.mpg.de}

\altaffiltext{1}{Universit\"ats-Sternwarte M\"unchen, Scheinerstra\ss e 1, D-81679 M\"unchen, Germany}
\altaffiltext{2}{Max-Planck-Institut f\"ur extraterrestrische Physik, Postfach 1312, Giessenbachstr., D-85741 Garching, Germany}
\altaffiltext{3}{Max-Planck-Fellow}

\begin{abstract}
We investigate the origin and fate of the recently discovered 
gas cloud G2 close to the Galactic Center.
Our hydrodynamical simulations focussing on the dynamical evolution of the cloud in combination
with currently available observations favor two scenarios: a {\it Compact Cloud}
which started around the year 1995 and a {\it Spherical Shell} of gas, with an 
apocenter distance within the disk(s) of young stars and a radius of a few times the size of the 
{\it Compact Cloud}. 
The former is able to explain the detected signal of G2 in the position-velocity diagram of 
the Br$\gamma$ emission of the year 2008.5 and 2011.5 data. 
The latter can account for both, G2's signal as well as the fainter extended tail-like structure G2t
seen at larger distances from the black hole and smaller velocities. 
In contrast, gas stripped from a compact cloud by hydrodynamical interactions is not able to explain the location of the detected 
G2t emission in the observed position-velocity diagrams. 
This favors the {\it Spherical Shell Scenario} and 
might be a severe problem for the {\it Compact Cloud} as well as the so-called {\it Compact Source Scenario}.
From these first idealized simulations we expect a roughly constant feeding of the supermassive black hole 
through a nozzle-like structure over a long period, 
starting shortly after the closest approach in 2013.51 for the {\it Compact Cloud}. 
If the matter accretes in the hot accretion mode, we do not expect a significant boost of 
the current activity of Sgr\,A* for the {\it Compact Cloud} model, but a boost of the average
infrared and X-ray luminosity by roughly a factor of 80 for the {\it Spherical Shell} scenario with
order of magnitude variations on a timescale of a few months.
Assuming that a part of the gas is accreted in cold disk mode, even higher boost factors can be reached.
The near-future evolution of the cloud will be a sensitive probe of the conditions of the gas distribution
in the milli-parsec environment of the massive black hole in the Galactic Center. 
\end{abstract}




\section{Introduction}
\label{sec:intro}

The Galactic Center harbors a nuclear supermassive black hole (SMBH). 
Due to its proximity, its mass of $4.31\times 10^6\,M_{\odot}$
\citep{Gillessen_09,Ghez_08,Genzel_10} 
has been derived from direct long-term ($\approx$20 years) monitoring of stellar orbits in the 
sub-parsec-scale vicinity of the central compact nonthermal radio source Sgr\,A* -- thought to be the radiative 
counterpart of the central SMBH. 
The uncertainties in the mass-estimate of the order of 10\% are mostly due to the distance 
measurement.  
The parsec-scale environment of the center is comprised of seven main constituents \citep[e.~g.~][]{Melia_01, Genzel_10}:
an atmosphere of diffuse hot gas (e.~g.~\citealp{Baganoff_03},\citealp{Yuan_03}, see also Sect.~\ref{sec:hot_atm}), 
a cluster of evolved stars \citep{Genzel_03a}, 
one or two warped disks of young stars ranging between 0.04\,pc and 0.5\,pc \citep{Genzel_03a,Levin_03,Bartko_10,Lu_09,Alig_11},
the B-stars in the central half-parsec \citep{Gillessen_09},  
spiral shaped thermal gas streams in the inner 2\,pc, which reaches down to 0.17\,pc 
(called Sgr\,A West, \citealp{Ekers_71,Downes_71,Ekers_83,Yusef_93,Zhao_09}),
a molecular dusty ring ranging from 2\,pc to 5\,pc \citep{Genzel_85, Montero_09} and a shell-like, elliptical non-thermal structure
with a size of roughly 6 $\times$ 8\,pc (called Sgr\,A East, \citealp{Ekers_71,Downes_71,Ekers_75,Maeda_02}). 
For a more detailed review of the Galactic Center environment we refer to \citet{Genzel_10}. 

\begin{figure*}
\epsscale{1.0}
\includegraphics[width=0.95\textwidth]{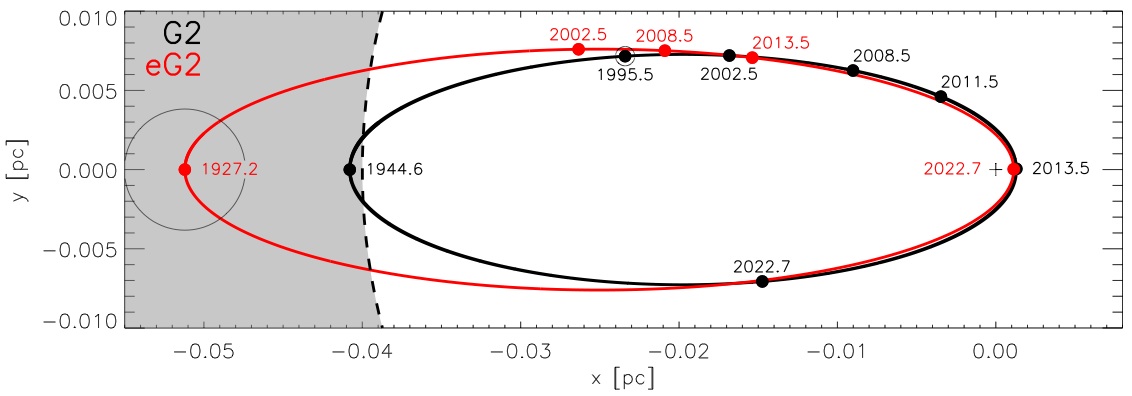}
\caption{Orbits of our two scenarios: The black line represents the orbit of G2, as determined in 
\citet{Gillessen_12}, representative for the {\it Compact Cloud Scenario} and the red line
denotes the orbit of eG2 as determined in this work, representative for the {\it Spherical Shell Scenario}.
The location of the central black hole is denoted by the plus sign and the dashed line denotes the inner 
rim of the disk(s) of young stars. The sizes of the shell and the cloud are shown by the thin black lines.}
\label{fig:orbits}
\end{figure*}

Recently, a gas cloud has been detected on its way towards the Galactic Center \citep{Gillessen_12}.
It consists of a two-component system: (i) the high surface brightness compact component G2 and (ii) a lower surface
brightness cone-like structure, which we will refer to as G2t from now on, as it was originally described as the {\it tail}
of G2 \citep{Gillessen_12}. The whole complex (G2 plus G2t) will be called eG2, the {\it extended G2} cloud complex, 
in order to avoid confusion.  
Traced back on archival data, \citet{Gillessen_12} were able to determine an accurate 
orbit for G2 (Fig.~\ref{fig:orbits}), which has a very high eccentricity ($e=0.9384$). 
It will reach its closest approach of only $4 \times 10^{15}\,$cm to the SMBH in 2013.51 
(corresponding to 3100 Schwarzschild radii).
Clear evidence for ongoing tidal velocity shearing and stretching of G2 within the last couple of years has been detected,
confirming its extended nature. 
This provides an interesting challenge for hydrodynamical simulations that aim at reproducing the currently visible tidal shearing
and that can be used to predict the future tidal evolution over the next couple of years.
G2 comprises of a multi-phase medium with the bulk of its mass being ionized gas ($M_{\mathrm{G2}}\approx 1.7\times 10^{28}\,f_{\mathrm{V}}\,$g, 
with $f_{\mathrm{V}}$ being G2's volume filling factor, which we assume to be unity throughout this paper). 
Ionization equilibrium with the radiation field of the nearby massive stars is expected to 
keep the temperature of this component at around $10^4$\,K. 
Additionally, it contains a small amount of warm dust with a temperature of $\approx 550\,$K. Only very little is known about
G2t, but its mass could be substantially larger than the mass estimate for G2 itself. 

Several questions are of immediate theoretical interest:
\begin{itemize}
  \item Which physical process formed the cloud?
  \item What determines the physical cloud properties (mass, size, thermodynamic state) and the dynamics?
  \item Can it be used to probe the direct vicinity of Sgr\,A*?
  \item What is the fate of the cloud in the next years?
  \item How does it influence the activity of Sgr\,A*? 
\end{itemize}

In \citet{Burkert_12} (paper\,I) we mainly tackled the first three of these questions mostly with the help of analytical considerations.
We especially concentrated on possible formation scenarios and evolutionary paths of G2. 
Two basic models proved to be the most successful:
\begin{itemize}
  \item {\bf The Compact Cloud Scenario (CC)}\\
        A cloud with uniform density structure can explain the evolution of the observed position-velocity (PV) diagrams of G2.
  \item {\bf The Spherical Shell Scenario (SS)}\\
        A tidally disrupted, initially spherical shell of gas with a much larger radius and mass than G2 can account for both, 
        the G2 and G2t component in the position-velocity diagram. 
\end{itemize}
In paper\,I we also found that the cloud is a sensitive probe of the hot accretion zone of the SMBH and the 
currently available observations
of G2 are in good agreement with models of captured, shock-heated stellar winds. We were able to derive constraints on the mass
of G2, which is above the critical mass below which clouds would evaporate before reaching the presently observed radius, and
which is small enough for the sound crossing timescale at apocenter to be of order its orbital period. 
Analytical estimates of various other time scales showed that many physical processes might contribute to the evolution of the cloud.
Therefore, detailed numerical simulations are needed, which is the topic of this publication.
The wealth of available observational data (now and in future) as well as its proximity make this event an ideal 
tool to study gas accretion physics in the vicinity of SMBHs. It will also give us very valuable insight into the 
evolution and the activation of galactic nuclei in general.

An alternative model to our {\it Compact or Spherical Cloud Scenario} -- called the 
{\it Compact Source Scenario} in paper\,I -- 
was proposed by \citet{Murray_Clay_11} (see also \citealp{Miralda_12}): a dense, proto-planetary disk, 
which is bound to its parent low-mass star.
The star is assumed to be scattered roughly 100 years ago from the observed disk(s) of young stars. Far away from Sgr\,A*, the 
protostellar gas disk remains bound to the star and resists tidal stretching. When approaching the SMBH, it looses mass through 
photo-evaporation due to the extreme flux of ionizing and FUV photons in the Galactic Center and due to tidal stretching. 
\citet{Murray_Clay_11} analytically estimate that this results in a cloud with similar properties as observed. 

The main idea of this paper is to present a first set of idealized hydrodynamical simulations 
and to test the influence 
of the hot atmosphere on the late-time evolution of such a cloud 
for our two best-fit cloud infall models as already described in paper\,I, namely
the {\it Compact Cloud Scenario} and the {\it Spherical Shell Scenario}. 
We discuss simple test particle simulations in Sect.~\ref{sec:testpart} before we describe the model
we employ for the Galactic Center environment in Sect.~\ref{sec:mod_atmosphere} and review
some of the complicated physical processes in Sect.~\ref{sec:involv_phys}.
The results of our hydrodynamical simulations are presented in Sect.~\ref{sec:hydro_results}, followed by a 
discussion (Sect.~\ref{sec:discussion}) and finally the conclusions (Sect.~\ref{sec:conclusions}).
We assume a Galactic Center distance of 8.33\,kpc, 
where $1\arcsec=40\,\mathrm{mpc}=1.25\times 10^{17}\,\mathrm{cm}$ \citep{Gillessen_09}.

\section{Test particle simulations}
\label{sec:testpart}

In order to find initial conditions for our hydrodynamical studies, which are able to reproduce 
current observations, we calculate a 
grid of test particle simulations in two dimensions, 
varying a large number of the parameters, as described below. This is done 
for both models of origin as discussed in Sect.~\ref{sec:testpart_cc} and \ref{sec:testpart_ss}.
Such kind of simulations are very fast and hence a large number of test calculations can be done on a short timescale 
(approximately 16\,000 simulations for this paper), allowing for a huge scan of the involved parameters.
The test particle simulations are run with the N-body code REBOUND \citep{Rein_12} and 100 test particles are distributed
along the boundary of the cloud or ring, restricted to the orbital plane. The particles are then evolved 
in the gravitational potential of the black hole and describe the evolution of the shape of the cloud when
tidal forces completely dominate. The best fit model is found by comparing the 
simulated to the observed position velocity diagrams.   
We use these N-body simulations to roughly constrain the formation time and initial conditions of G2 
and eG2, respectively.
The best fitting initial conditions are then used to study the detailed hydrodynamical evolution 
(Sect.~\ref{sec:hydro_results}).

\subsection{Compact Cloud Scenario}
\label{sec:testpart_cc}

\begin{figure}[b]
\epsscale{1.0}
\includegraphics[width=0.95\columnwidth]{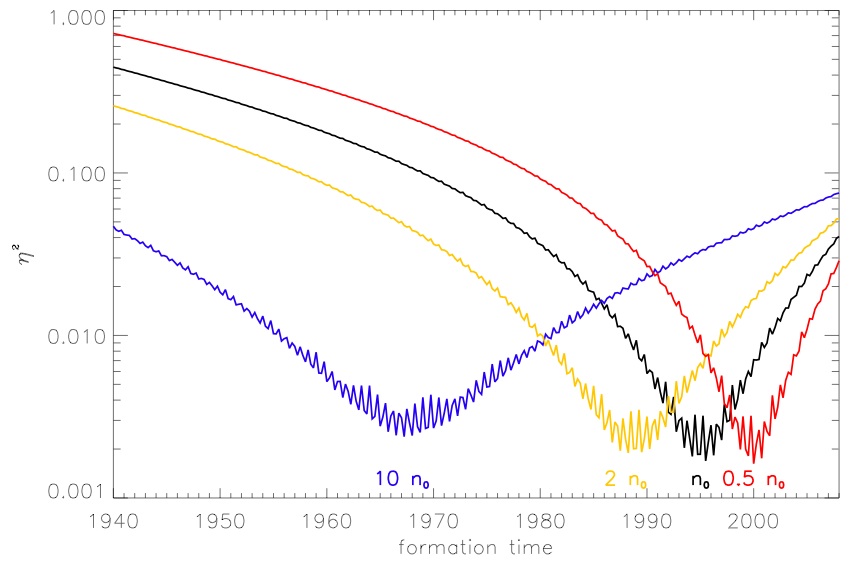}
\caption{Possible formation times of the compact cloud models as inferred from test particle simulations for
various densities of the atmosphere. The black line corresponds to our standard atmosphere, whereas the 
colored lines refer to multiples of the density as indicated in the plot. 
The $\eta^2$ value quantifies the goodness of fit with
the observed position-velocity diagrams in 2008.5 and 2011.5 (see Sect.~\ref{sec:testpart_cc}).}
\label{fig:CC_study}
\end{figure}

For the compact cloud scenario, we assume that the cloud started {\it in situ} in pressure 
equilibrium with the surrounding hot atmosphere on the observed orbit of G2 and that G2's mass
is known\footnote{The observationally determined mass actually depends on the unknown volume filling
factor, which we assume to be one here.}. Then, the initial radius and density of the cloud are 
fixed along the orbit and we only vary the date of birth of the cloud. 
As estimated in paper\,I, hydrodynamical effects -- especially ram pressure --  play a minor role
in the early evolution of the cloud in the compact cloud scenario. This means that we can directly 
compare the resulting distribution in the position-velocity diagram of the year 2008.5 and 2011.5 with 
the observations and can expect to get a reasonably well determined initial condition. 
The goodness of fit is judged from the position and spread of the test particles in the PV diagram. We 
compare the minimum and maximum values of the projected distance ($d_{\mathrm{min,max}}^{\mathrm{sim}}$) 
as well as the line of sight velocity ($v_{\mathrm{min,max}}^{\mathrm{sim}}$) of the simulated distribution with 
the respective extrema of the observed 
PV diagram ($v_{\mathrm{min,max}}^{\mathrm{obs}},d_{\mathrm{min,max}}^{\mathrm{obs}}$). 
For the latter we chose a representative contour line (0.8 times the maximum value of the observed PV diagram, 
shown by the two plus signs in the left two panels of Fig.~\ref{fig:pvobs}). 
The quadratic sum of these relative deviations of the 2008.5 and 2011.5 observations with the respective 
simulated snapshots are summarized in an $\eta^2$-value, used to judge the goodness of the fit:

\begin{eqnarray}
\label{equ:eta}
\eta^2 = \sum_{2008.5,2011.5}\,\left(\frac{v_{\mathrm{min,max}}^{\mathrm{sim}}-v_{\mathrm{min,max}}^{\mathrm{obs}}}{v_{\mathrm{min,max}}^{\mathrm{obs}}}\right)^2 \nonumber\\
                           + \left(\frac{d_{\mathrm{min,max}}^{\mathrm{sim}}-d_{\mathrm{min,max}}^{\mathrm{obs}}}{d_{\mathrm{min,max}}^{\mathrm{obs}}}\right)^2
\end{eqnarray}

Fig.~\ref{fig:CC_study} displays the result of this study for our assumed standard
atmosphere (see Sect.~\ref{sec:hot_atm}). It shows a clear $\eta^2$ minimum around the year 1995.5. 
A study of different density normalizations of the atmosphere (Fig.~\ref{fig:CC_study})
shows that the preferred origin of the cloud depends sensitively on the density distribution of the atmosphere.
Increasing the atmospheric density by a factor of two 
shifts the starting time to the year 1988, whereas a factor of ten would necessitate a starting time 
of 1967 and a decrease of the density distribution by a factor of two yields a starting date of roughly
2000. This can be understood as the cloud sizes depend on the atmospheric pressure, when we assume 
pressure equilibrium between the cloud and the atmosphere. 
However, the values also depend on the choice of the limiting contour for the fitting procedure.
Therefore, these studies only serve as rough estimates for possible origin dates of the cloud. 
To determine and describe the future hydrodynamical evolution, we will start the cloud in 1995.5 
(Sect.~\ref{sec:denevol_cc}).

\subsection{Spherical Shell Scenario}
\label{sec:testpart_ss}

\begin{figure*}
\epsscale{1.0}
\includegraphics[width=0.95\textwidth]{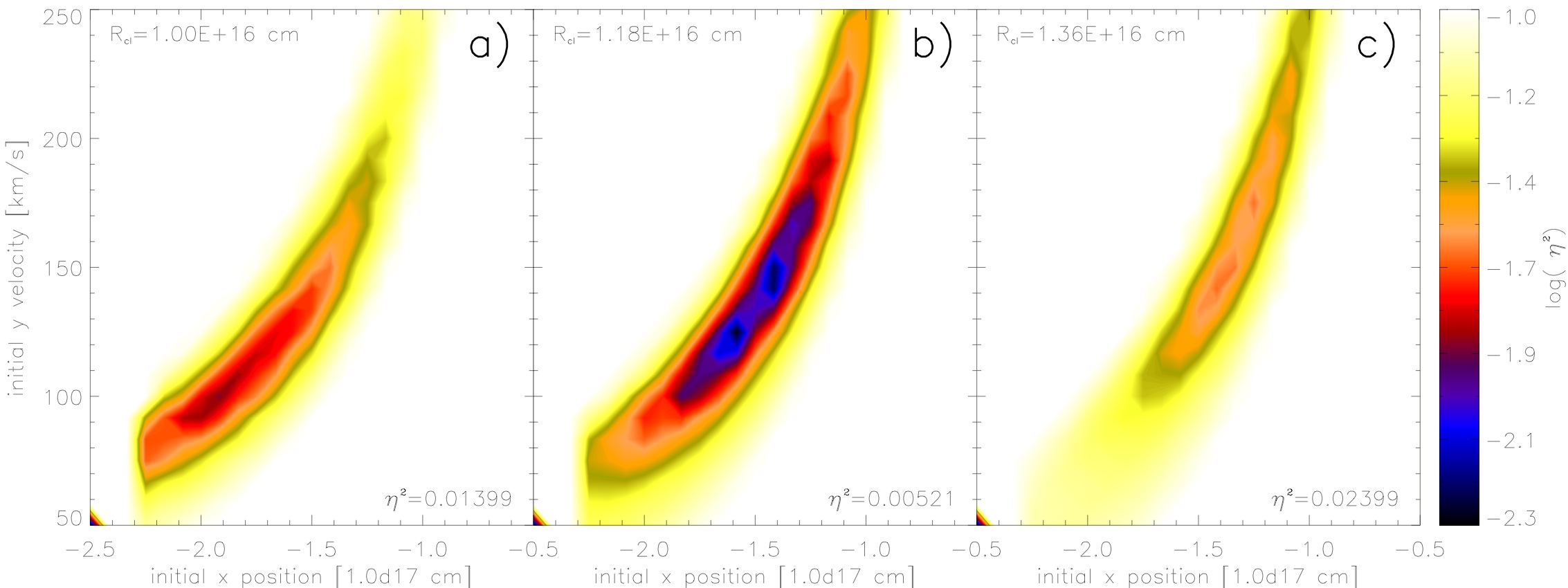}
\caption{Distribution of $\eta^2$-values (see Sect.~\ref{sec:testpart_cc}) of the spherical shell scenario as inferred from test particle simulations in the 
initial x-position vs. initial y-velocity plane. We assume the ring starts at arbitrary time at the apocenter,
which is located on the negative y-axis. The three different panels refer to three exemplary 
initial ring radii: $1.00\times 10^{16}$\,cm (a), $1.18\times 10^{16}$\,cm (b) and $1.36\times 10^{16}$\,cm (c).
The $\eta^2$ value quantifies the proximity of the match with
the observed position-velocity diagrams in 2008.5 and 2011.5 (see text for details of the fitting procedure).}
\label{fig:SS_study}
\end{figure*}

As shown in paper\,I, the observations can also be interpreted in such a way that
G2 and G2t are parts of a larger cloud complex (eG2), which is already partly disrupted by tidal forces
in the 2008.5 and 2011.5 observations. 
As these shells are typically a factor of five to ten larger than the compact clouds mentioned above, ram pressure effects 
cannot be neglected anymore. In order to find reasonable starting parameters for this model as well, we 
take the ram pressure effects roughly into account by offsetting the comparison values from the observed contour  
(compare to blue plus signs in panels\,c,d of Fig.~\ref{fig:pvobs}).
This offset is determined by comparing a number of hydrodynamical simulations with test particle simulations.
Thought to be the result of emission of stellar material, we assume that this shell started close to the apocenter of 
its elliptical orbit, which lies well within the disk(s) of young stars (see Fig.~\ref{fig:orbits}), 
making it the most likely origin of the cloud. 
For this model, the total mass of the cloud is not fixed, as the mass of G2t cannot be constrained observationally 
very well at present. In the subsequent hydrodynamical modelling we then choose the density of the shell to reach approximate 
pressure equilibrium with the surrounding atmosphere. 
Again, we are only interested in a rough match of the hydrodynamical realization with the observed PV-diagram in order to 
investigate the basic evolution of such a model. In this case, every time step is compared to the 2008.5 PV diagram and the stage
three years later with the 2011.5 observation in order to find the best matching timestep, 
as the date of birth is not fixed in this model.
The result of this fitting procedure with a grid of $25^3$ test particle simulations is summarized in Fig.~\ref{fig:SS_study}.
Shown is the distribution of $\eta^2$ values determined as described in Sect.~\ref{sec:testpart_cc} and 
equation \ref{equ:eta} in a plane spun by the initial position on the
negative x-axis and the initial velocity in y-direction. The panels display the distribution for three examples
of the initial outer ring radius: $1.00\times 10^{16}$cm (panel\,a), $1.36\times 10^{16}$cm (panel\,c) and our 
best fit value of $1.18\times 10^{16}$cm (panel\,b). 
The minimum $\eta^2$ valley is quite shallow and we take the global minimum of our simulated grid, which is a model
starting at apocenter in the year 1927.2 at a distance from the black hole of $1.58\times 10^{17}\,$cm and an initial
orbital velocity of $125\,\mathrm{km}\,\mathrm{s}^{-1}$.

\section{Modelling the Galactic Center environment}
\label{sec:mod_atmosphere}

\subsection{The hot atmosphere}
\label{sec:hot_atm}

X-ray observations of hot gas emission near Sgr\,A* \citep{Baganoff_03} are attributed to the atmosphere close to the 
SMBH, which can be explained as gas being shock-heated by the interaction of strong stellar winds of the 
surrounding massive stars 
\citep[e.~g.~][]{Krabbe_91,Melia_92,Baganoff_03,Muno_04}.
The observed very low bolometric luminosity of the order 
of $3\times 10^{-9}\,L_{\mathrm{Edd}}$ together with a Bondi accretion rate of $10^{-5}\,M_{\odot}\mathrm{yr}^{-1}$ 
\citep{Yuan_03} disfavor standard accretion disk theory \citep{Shakura_73}.
They are rather consistent with 
so-called {\it Radiatively Inefficient Accretion Flows} (RIAFs), in which only a small fraction of 
the gravitational energy is radiated. 
Following \citet{Gillessen_12}, we adopt the
ADAF ({\it Advection-Dominated Accretion Flow}, e.~g.~\citealp{Ichimaru_77, Abramowicz_95,Narayan_94}) model of 
\citet{Yuan_03}, which is an analytical approximation for such a RIAF solution, in 
concordance with available Chandra observations \citep{Baganoff_03} and
rotation measure data derived from radio observations \citep{Bower_03}.
In the idealized simulations presented in this paper, 
we neglect the rotation and the actual dynamics of the accretion flow
and rescale the normalization of the temperature distribution accordingly, in 
order to yield hydrostatic equilibrium.
This yields the following number density distribution $n_{\mathrm{at}}$ and
temperature distribution $T_{\mathrm{at}}$ of the hot atmosphere:

\begin{eqnarray}
\label{equ:numdens_at}
n_{\mathrm{at}} = 930\,\mathrm{cm}^{-3}\,f_{\mathrm{hot}}\,\left(\frac{1.4\times 10^4 R_{\mathrm{S}}}{r}\right)^{\alpha} \\
\label{equ:temp_at}
T_{\mathrm{at}} = 1.2\times 10^8\,\mathrm{K}\,\left(\frac{1.4\times 10^4 R_{\mathrm{S}}}{r}\right)^{\beta}
\end{eqnarray}

where $r$ is the distance to the black hole, 
$R_{\mathrm{S}}$ is its Schwarzschild radius, $\alpha \approx \beta \approx 1$ and $f_{\mathrm{hot}}\approx 1$ is a factor taking the
uncertainty of the model into account.
A mean molecular weight of $\mu=0.6139$ has been assumed, typical for a gas with solar metallicity.

However, as the entropy $S = T / \rho^{2/3}$ decreases outwards, this is not a stable 
equilibrium, but unstable to convection (see discussion in paper\,I), which
we confirmed with the help of test simulations: small perturbations start growing 
at the inner boundary surrounding the black hole, which quickly leads to the formation of 
unphysically high pressure bubbles.
Due to the unstable nature of the atmosphere we do expect an inhomogeneous density distribution.
However, setting up a realistic RIAF solution is a complicated task in itself 
(e.~g.~\citealp{Stone_99}, \citealp{Stone_01} and \citealp{Igumenshchev_03}) and beyond the scope 
of this first idealized study presented in this article 
as it would require the inclusion of many more physical processes, like e.~g.~magnetic fields 
(see Sect.~\ref{sec:magfields} for a discussion of possible effects), thermal conduction, etc.
Some of these effects are already taken into account (not completely self-consistently) in the
models of \citet{Yuan_03}.
We are rather interested in the interaction of the cloud
with a smooth atmosphere, representing an average distribution in concordance with
available observations.
To this end, we artificially stabilize it and keep the atmosphere's density and temperature distribution
fixed with time. 
This is done by additionally evolving a passive tracer field ($0 \leq tr \leq 1$), which allows us to distinguish 
between those parts of the atmosphere which have interacted with the cloud ($tr \geq 10^{-4}$, 
see Sect.~\ref{sec:tracer_eff}) from those which changed
due to the atmosphere's inherent instability ($tr < 10^{-4}$). Those cells fulfilling the latter criterion 
are reset to the values expected in hydrostatic equilibrium. 
We discuss numerical issues and proof that this 
is a sensible assumption in Sect.~\ref{sec:tracer_eff} below.

\subsection{The initial conditions of the cloud models and the numerical treatment}
\label{sec:inicon}

We use the two best-fit models presented in Sect.~\ref{sec:testpart} as basis for our hydrodynamical 
exploration of the origin and fate of the cloud complex eG2. 
In these first simulations, we generally set the adiabatic index $\Gamma=1$, but also 
present the comparison to a purely adiabatic simulation.
This seems to be a reasonable assumption, as the temperature structure of the atmosphere is supposed to
be given by adiabatic heating of the accretion flow itself and stellar processes (see Sect.~\ref{sec:hot_atm})
and the temperature of the cloud material is expected to be set by photoionization equilibrium
in the radiation field of the surrounding stars \citep{Gillessen_12} during the early evolution 
of the cloud.
In order to explain the observed position-velocity diagram, 
two basic scenarios have been identified in paper\,I. 
Both are based on the simple assumption that the cloud started 
with spherical shape and in pressure equilibrium with the atmosphere on an orbit compatible with 
the observed positions on the sky. In the first set of models we follow the interpretation of 
\citet{Gillessen_12}, who discern the observations into {\it cloud head} and {\it tail} emission.
Our second set of models interprets the observations as being the result of the disruption of a
larger cloud, which -- for concordance with observations -- necessitates the cloud to be a spherical shell.

Our two-dimensional simulations are done within the orbital plane of the cloud, with the major axis parallel to the
x-axis and the apocenter of the orbit on the negative x-axis of our fixed Cartesian coordinate system 
with a spatial resolution of $7\times 10^{13}$cm. The computational grid of our standard model ranges from  
$-1.3\times 10^{17}$cm to $1.2\times 10^{16}$cm in x-direction
and $-6.2\times 10^{16}$cm to $2.5\times 10^{16}$cm in y-direction.
The cloud is on a clockwise orbit with the major axis parallel to the x-axis and the pericenter 
of the orbit on the positive x-axis.
The black hole is located at the origin of 
our coordinate system.
As discussed above we use a passive tracer field in order to trace the evolution of the cloud 
(and to stabilize the atmosphere)
and initially assign the cloud a value of 1 and the atmosphere 0 and let it evolve.  

The hydrodynamical equations are integrated with
{\sc PLUTO}, version 3.1.1 \citep{Mignone_07}.
For all simulations shown in this article, we use the two-shock Riemann solver,  
do a parabolic interpolation and employ the third order Runge-Kutta time integration scheme.
A cartesian two-dimensional grid is used and 
the boundary conditions in all directions are set to the values expected
for hydrostatic equilibrium of the hot atmosphere, 
enabling outflow but no inflow. We introduce an additional spherical 
boundary surrounding the central black hole. All gas flowing into this part of the integration 
domain is removed and counted as being accreted. All values inside this boundary are set to the 
atmospherical values, in order to mimick the pressure exerted by this region, which is 
necessitated by the steep pressure gradient.

We neglect thermal conduction, magnetic fields as well as feedback from the central source for the sake of simplicity. 
A summary of the used and varied parameters is given in Table\,\ref{tab:obsparam} and Table\,\ref{tab:simparam}, 
respectively.


\begin{table}[b]
\begin{center}
\caption{Observed and calculated parameters of the cloud G2 and the Galactic Center region.\label{tab:obsparam}}
\begin{tabular}{lll}
\tableline\tableline
Parameter & & Value \\
\tableline
mass of the SMBH & $M_{\mathrm{BH}}$ & $4.31\times 10^6\,M_{\odot}$ \\
Galactic Center distance & $R_0$ & $8.33$\,kpc \\
mass of the cloud G2\tablenotemark{a} & $m_{\mathrm{cloud}}$ & $1.7\times 10^{28}$\,g\\
pericenter distance & $r_{\mathrm{peri}}$ & $4\times 10^{15}$\,cm\\
time of closest approach & $t_{\mathrm{peri}}$ & $2013.51$\\
orbit eccentricity & $\epsilon$ & 0.9384 \\
semi-major axis & $a_{\mathrm{semi}}$ & $6.49\times 10^{16}\,$cm \\
orbital period & $T_{\mathrm{orbit}}$ & $137.77$ \\
inclination of ascending node & $i$ & $106.55$ \\
position angle of ascending node & $\Omega$ & $101.5$ \\
longitude of pericenter & $\omega$ & $109.59$ \\
\tableline
\end{tabular}
\tablecomments{Parameters of G2 and the Galactic Center obtained or derived from observations in \citet{Gillessen_12}.
(a) Throughout the paper we assume a volume filling factor of one.}
\end{center}
\end{table}

\begin{table*}
\begin{center}
\caption{Parameters of the (hydrodynamical) simulations.\label{tab:simparam}}
\begin{tabular}{lllllllllllll}
\tableline\tableline
  & $\tau_{0}$\tablenotemark{a} &  $\rho_{\mathrm{cloud}}$\tablenotemark{b} & $R_{\mathrm{cloud}}$\tablenotemark{c} &
$x_{\mathrm{ini}}$\tablenotemark{d} & $y_{\mathrm{ini}}$\tablenotemark{e} & $v^x_{\mathrm{ini}}$\tablenotemark{f} &
$v^y_{\mathrm{ini}}$\tablenotemark{g} & $\Gamma$\tablenotemark{h} & $f_{\mathrm{hot}}$\tablenotemark{i} & 
$\Delta x$\tablenotemark{j} & $tr$\tablenotemark{k} \\
 & {\footnotesize yr AD} & {\footnotesize $10^{-19}\,\mathrm{g}\,\mathrm{cm}^{-3}$} & {\footnotesize $10^{15}\,\mathrm{cm}$} & 
{\footnotesize $10^{16}\,\mathrm{cm}$} & {\footnotesize $10^{16}\,\mathrm{cm}$} 
& {\footnotesize $\mathrm{km}\,\mathrm{s}^{-1}$} & {\footnotesize $\mathrm{km}\,\mathrm{s}^{-1}$} &  &  & {\footnotesize $10^{13}\,\mathrm{cm}$} &  \\
\tableline
CC01 & $1995.5$ & $6.21$   & $1.87$ & $-7.22$  & $ 2.21$ & $794.59$ & $48.45$ & $1$ & $1$ & $7.0$ & $10^{-4}$ \\
CC02 & $1944.6$ & $2.24$   & $2.63$ & $-12.59$ & $0.0$   & $0.0$    & $167.29$ & $1$ & $1$ & $7.0$ & $10^{-4}$  \\
CC03 & $1944.6$ & $223.64$ & $0.57$ & $-12.59$ & $0.0$   & $0.0$    & $167.29$ & $1$ & $1$ & $7.0$ & $10^{-4}$  \\
CC04 & $1995.5$ & $6.21$   & $1.87$ & $-7.22$  & $ 2.21$ & $794.59$ & $48.45$ & $5/3$ & $1$ & $7.0$ & $10^{-4}$  \\
CC05 & $2000.0$ & $4.33$ & $2.11$ & $-5.99$ & $2.24$ & $953.00$ & $-5.41$ & $1$ & $0.5$ & $7.0$ & $10^{-4}$  \\
CC06 & $1988.0$ & $8.57$ & $1.68$ & $-8.87$ & $2.03$ & $605.85$ & $98.85$ & $1$ & $2.0$ & $7.0$ & $10^{-4}$  \\
CC07 & $1995.5$ & $6.21$   & $1.87$ & $-7.22$  & $ 2.21$ & $794.59$ & $48.45$ & $1$ & $1$ & $14.0$ & $10^{-4}$  \\
CC08 & $1995.5$ & $6.21$   & $1.87$ & $-7.22$  & $ 2.21$ & $794.59$ & $48.45$ & $1$ & $1$ & $3.5$ & $10^{-4}$  \\
CC09 & $1995.5$ & $6.21$   & $1.87$ & $-7.22$  & $ 2.21$ & $794.59$ & $48.45$ & $1$ & $1$ & $7.0$ & $10^{-6}$ \\
CC10 & $1995.5$ & $6.21$   & $1.87$ & $-7.22$  & $ 2.21$ & $794.59$ & $48.45$ & $1$ & $1$ & $7.0$ & $10^{-2}$ \\
SS01  & $1927.2$ & $1.42$ & $11.80$\tablenotemark{l} & $-15.80$ & $0.0$ & $0.0$ & $125.00$ & $1$ & $1$ & $7.0$ & $10^{-4}$  \\
\tableline
\end{tabular}
\tablecomments{CC refers to simulations of the Compact Cloud scenario and SS to those of the Spherical Shell scenario. 
Simulations CC02 and CC03 are only discussed in paper\,I. (a) Start time of the simulation, (b) initial density of the cloud,
(c) initial radius of the cloud, (d) initial x-position of the cloud, (e) initial y-position of the cloud, (f) initial x-velocity of the cloud,
(g) initial y-velocity of the cloud, (h) adiabatic index, (i) uncertainty factor of the density distribution of the hot atmosphere, 
(j) cell size in x- and y-direction, (k) tracer threshold value, (l) the shell has a thickness of 25\% of this outer radius.}
\end{center}
\end{table*}


\section{Cloud physics}
\label{sec:involv_phys}

The components of the Galactic Center environment which have the largest impact on the gas cloud are the gravity of the central 
SMBH and the hydrodynamical interaction with the hot atmosphere. Therefore, we restrict our discussion to the physical 
effects related to these two components. 
The relevant physical processes of the cloud/atmosphere system are discussed in great detail in paper\,I. 
Here, we only briefly 
summarize those important for the interpretation of our hydrodynamical simulations and estimate the 
timescales of the various processes for different phases of the evolution of the cloud.

\subsection{Tidal deformation}

Tidal forces lead to a stretching in direction of the black hole and a squeezing perpendicular to it. 
As G2's orbit has a very high eccentricity, the cloud gets distributed along the orbital path.
The acceleration of this stretching is given by 

\begin{eqnarray}
|a_{\mathrm{tf}}| =  2 \, \Delta r \, G \, \frac{M_{\mathrm{BH}}}{r^3},
\end{eqnarray}

where $\Delta r$ refers to the width of the cloud in direction of the black hole, $G$ is 
the gravitational constant, $M_{\mathrm{BH}}$ is the mass and $r$ the distance to the SMBH.
Therefore, a change of the cloud elongation of the order the width of the cloud 
by tidal forces is expected on a timescale of:

\begin{eqnarray}
\tau_{\mathrm{tf}} \approx \sqrt{\frac{2\,\Delta r}{|a_{\mathrm{tf}}|}} \approx 
       \frac{r^{3/2}}{\sqrt{G\,M_{\mathrm{BH}}}} = 10.60\,\mathrm{yr}\,\left(\frac{r}{10^{16}\,\mathrm{cm}}\right)^{3/2},
\end{eqnarray}

which varies between 27.6\,yr at the 1995.5 position of the cloud and 0.5\,yr at the pericenter.

\subsection{Ram pressure compression and deceleration}

Ram pressure of the surrounding medium has two effects: (i) Locally, it changes the cloud structure by compressing
the upstream part of the cloud and (ii) globally it decelerates the cloud, leads to angular momentum exchange 
with the atmosphere and finally accretion of matter towards the center. 
In paper\,I we analytically estimate the loss of kinetic energy for our compact cloud model to be 
of the order of a few percent under the assumption of a cloud evolving in pressure equilibrium.
In realistic hydrodynamical simulations, the situation is much more complex as we will show in detail in 
Sect.\,\ref{sec:hydro_results}.

\subsection{Thermal conduction and evaporation}
\label{sec:therm_cond}

Thermal conduction leads to the evaporation of parts of the cloud. 
Following \citet{Cowie_77} we find an evaporation time scale of 

\begin{eqnarray}
  \label{equ:evaptime}
  \tau_{\mathrm{evap}} = 81\,\mathrm{yr}
                       \left(\frac{r}{10\,r_{\mathrm{peri}}} \right)^{15/8}\,
                       \left(\frac{R_{\mathrm{c}}}{10^{15}\,\mathrm{cm}}\right)^{13/8}\nonumber\\
                       \frac{M_{\mathrm{c}}}{1.7\times 10^{28}\,\mathrm{g}}
\end{eqnarray}

in the so-called saturated limit 
for a spherical cold gas cloud immersed in the hot atmosphere described in equation \ref{equ:numdens_at} and \ref{equ:temp_at}. 
Assuming that the cloud stays spherical and in pressure equilibrium along the whole orbit, the evaporation time scale 
changes between 97\,yr for the cloud at its initial distance in the year 1995.5 and 10\,yr when the cloud reaches the
pericenter. However, this neglects tidal effects and hydrodynamical instabilities, which finally lead to
the disruption of the cloud. Thereby, the cloud increases in size and the evaporation time scale might decrease. 
Hence, evaporation effects get important mostly during the late evolution of the cloud.  

As was shown in paper\,I and briefly discussed here, 
all of the mentioned physical processes are acting on similar timescales close to the orbital timescale of the cloud. 
This is especially the case for the late-time evolution and necessitates the use of numerical hydrodynamical 
simulations to determine the fate of the cloud.

\newpage

\section{Results of the hydrodynamical simulations}
\label{sec:hydro_results}

\subsection{The evolution of the density distribution in the Compact Cloud Scenario (CC01)}
\label{sec:denevol_cc}

\begin{figure*}
\epsscale{0.95}
\includegraphics[width=0.95\textwidth]{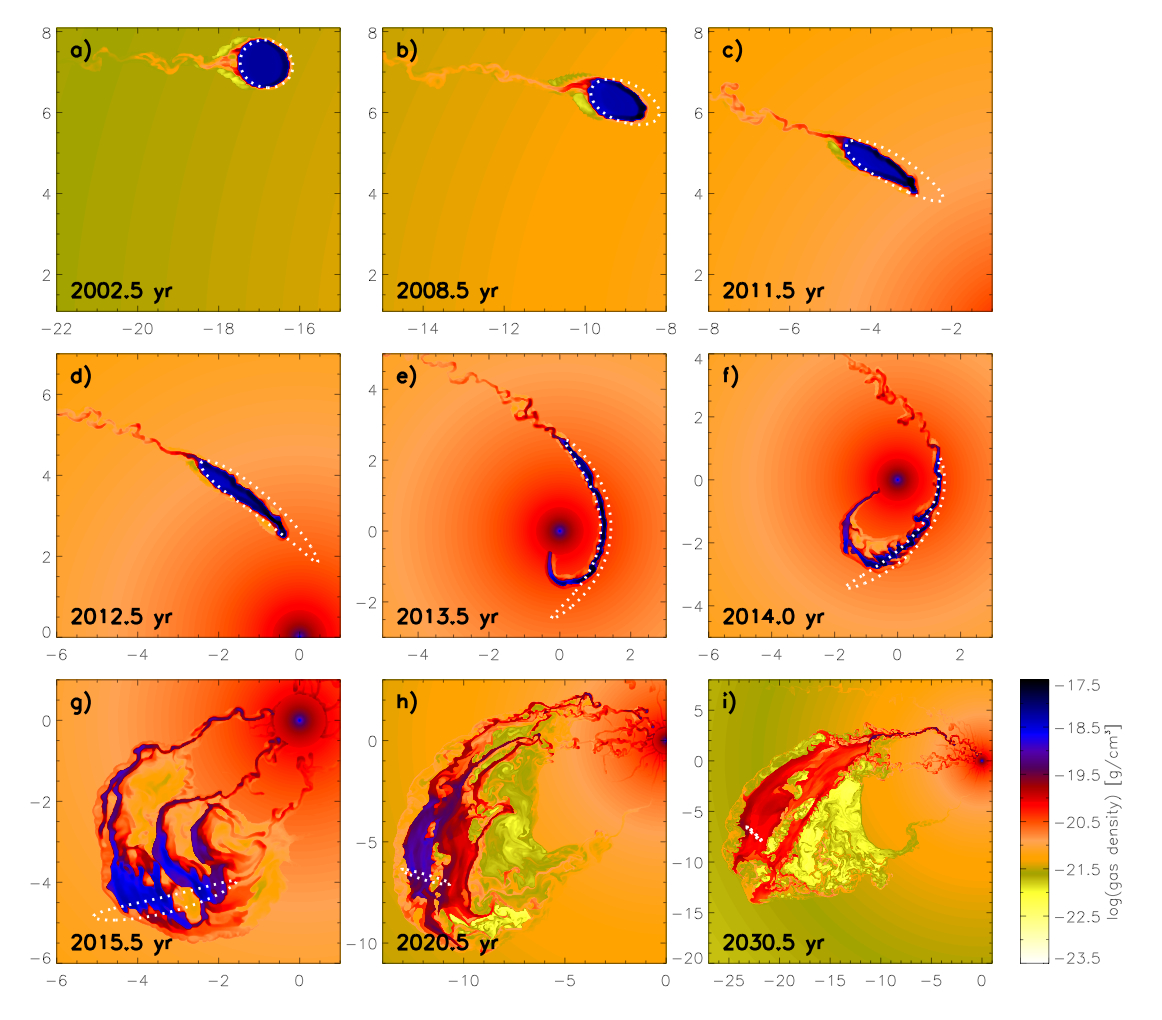}
\caption{Density evolution of the compact cloud model CC01. 
         Overlayed as dotted white lines are the positions of 
         test particles initially located at the cloud boundary.
         The axis labels are given in mpc (milli-parsec). Mind the different scalings of the panels.}
\label{fig:CC01_densevol}
\end{figure*}

The temporal evolution of the two-dimensional gas density distribution of model CC01 is 
shown in Fig.~\ref{fig:CC01_densevol}. Overlayed are the results of 
a simulation in which we placed test particles at the boundary of the initial cloud setup
and let them evolve under the influence of the gravitational potential alone. They serve
as an indicator for the differences in the evolution due to hydrodynamical effects.
We start with a phenomenological description of the cloud evolution and back our findings 
in the course of the paper with several analysis tools.
The initially spherical cloud begins its journey at the year 1995.5 position on G2's observed 
best-fit orbit. 
The early phase towards the pericenter which is reached in 2013.5 is 
dominated by tidal forces (panels\,a to e). 
Due to the highly eccentric orbit, the cloud gets stretched along the orbital direction. 
At the same time, the surrounding density and the increase in orbital velocity lead 
to an increase of ram pressure forces, resulting in 
a compression of the cloud head which can be inferred from the deviation of the density distribution 
from the test particle model (panel\,d). 
The shear flow at the boundary of the cloud leads to stripping of gas and the formation
of a long turbulent wake, which develops typical rolls of the Kelvin-Helmholtz instability (KHI, panel\,e). 
At the time of the pericenter passage, the cloud has developed into a long spaghetti-like 
filament more or less aligned with the test particle orbit of the cloud. 
This elongation of the cloud in combination with the large velocity and density difference 
(along G2's orbit) between the 
cloud and the atmosphere causes a rapid growth of the KHI, well visible by the meandering shape of
the cloud (panel\,f). In the subsequent evolutionary phase, a number of effects 
lead to the buckling of the cloud in direction of orbital motion (panel\,g):
First, the pure gravitational effect, leading to a deceleration of the front  
while the back still has a high velocity. This is the dominant effect, 
as the horizontal width of the test particle distribution almost coincides with the horizontal
width of the gas cloud.  
Second, this is the region of the 
orbit, where the atmosphere is the densest. Hence, strong ram pressure forces shape the 
head of the cloud and lead to additional compression along the orbit and elongation 
perpendicular to the orbital motion. Increasing the cross section, ram pressure forces 
get more and more important. A boundary layer builds
up at the front of the cloud, which looses angular momentum and cloud material is able to
accrete towards the center, forming a number of filaments pointing towards Sgr\,A* (panel\,g). Their 
number and location is dictated by the earlier growth of the Kelvin-Helmholtz instability 
mainly during the pericenter passage. In the course of the following evolution, the rear filaments
-- being in the {\it shadow} of the leading one -- merge with the one in front and 
form a nozzle-like structure of matter streaming towards Sgr\,A* and causing an increased 
accretion rate (Sect.~\ref{sec:accretion}, panel\,h).    
The later evolution of the cloud is dominated by hydrodynamical effects in our simulations: 
The large effective area of the cloud makes it susceptible to 
hydrodynamical instabilities and ram pressure. 
The phase during and shortly after the pericenter passage is also the time when for the 
first time, the cloud is able to mix with the atmosphere. Due to the large temperature
difference between cloud and atmosphere and the moderate density contrast in this region,
this mixing is able to increase the cloud temperature to approximately $10^5-10^6$\,K. 
At the end of this mixing process, about 1\% of the cloud is made up of atmospheric gas.

Whereas hydrodynamical instabilities shape the outer region of the cloud, 
ram pressure interaction with the atmosphere
enables further deceleration and accretion of gas from the boundary layer.
However, it is very difficult to make detailed predictions of the late evolution of the 
gas with the help of our idealized simulations, as by then, several other effects become
important. Thermal conduction will be able to act on the much increased surface area of 
the cloud and will evaporate it outside-in and the detailed thermodynamical treatment
might as well significantly change the morphology and accretion behavior in this 
evolutionary phase (see discussion in \ref{sec:eos}).

\begin{figure*}
\epsscale{0.95}
\includegraphics[width=0.95\textwidth]{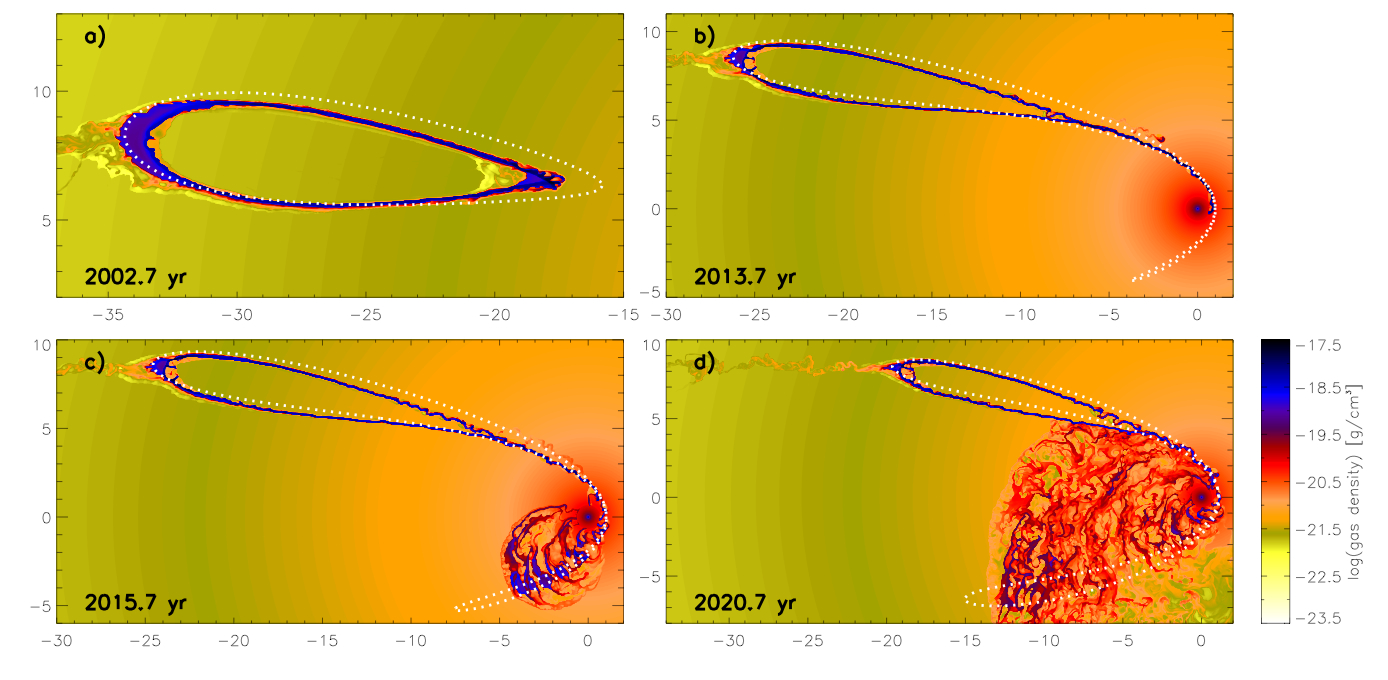}
\caption{Density evolution of the spherical shell model SS01. 
         Overlayed as dotted white lines are the positions of 
         test particles initially located at the outer ring boundary.
         The axis labels are given in mpc (milli-parsec). Mind the different scalings of the panels.}
\label{fig:SS01_densevol}
\end{figure*}

\subsection{The evolution of the density distribution in the Spherical Shell Scenario (SS01)}
\label{sec:denevol_ss}

Density snapshots showing the evolution of the spherical shell scenario (simulation SS01) are displayed in 
Fig.~\ref{fig:SS01_densevol}. The basic physical processes acting on the orbit until the 
pericenter passage are very similar to the evolution of the compact cloud.
Firstly, the ring squeezes into a drop-shaped morphology due to the interaction in the 
gravitational potential of the black hole (panel\,a). 
Secondly, the hydrodynamical interaction with the 
hot atmosphere leads to the formation of a turbulent wake behind the shell and 
at the ring border, KHI start to grow (panel\,b).
In a three-dimensional simulation, the structure would correspond to a hollow conical morphology, 
with one major focal point at the very front of the cloud. 
This is in good agreement with what is seen in observations: The focal point then 
corresponds to the actual cloud G2 and the conical structure is the diffuse tail G2t, which
indeed shows signs of a conical shape. In reality, the conical appearance is irregular and not continuous as 
the spherical shell is expected to have an inhomogeneous density distribution initially. 
At pericenter, the front part of the ring -- which we interpret as the cloud G2 -- has already 
collapsed into a thin, spaghetti-like
shape. In contrast to our compact cloud model (see Sect.~\ref{sec:denevol_cc}), 
the thin filament shows already the typical disturbances expected for the KHI 
and hence a larger effective cross section compared to the compact cloud (panel\,b). 
Being very close to the center of the hot atmosphere, the steep density gradient causes a rapid 
increase of the ram pressure, decelerating the front part of the ring.  
The KHI disturbances already present before pericenter passage grow on short timescale 
and finally lead to a fast disruption of the ring (panel\,c). 
It efficiently exchanges angular momentum 
with the dense inner atmosphere and a filamentary disk-like structure is able to form 
(panel\,d). 
This leads to a much higher mass accretion rate through the inner boundary (see Sect.~\ref{sec:accretion}).
We refer to Sect.~\ref{sec:discussion} for a critical discussion of the influence of numerical effects as 
well as our idealized treatment of the atmosphere.

\subsection{Angular momentum redistribution and mass accretion towards the center}
\label{sec:accretion}

\begin{figure}[b]
\epsscale{1.0}
\includegraphics[width=0.95\columnwidth]{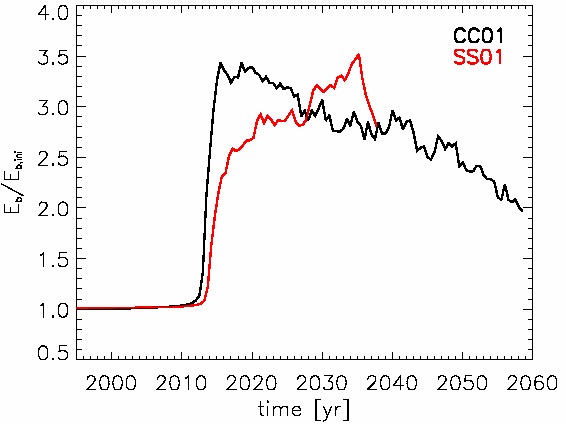}
\caption{Binding energy of the cloud/ring in units of the initial binding energy.}
\label{fig:bind_energy}
\end{figure}

The temporal evolution of the binding energy of the cloud (selected with the usual tracer threshold limit)
is shown in Fig.~\ref{fig:bind_energy} for the two models CC01 and SS01. 
As already discussed in paper\,I, the loss of kinetic 
energy due to ram-pressure in the early evolutionary phase is minor and the binding energy stays more or
less constant until the first part of the cloud reaches the dense inner part of the hot atmosphere around the year 2012/2013,
where we see a strong increase of ram pressure deceleration. 
As a consequence, the cloud gets bound more strongly to the SMBH. Ram pressure forces 
increase with the effective cross section of the cloud, they rise sharply shortly after the 
pericenter passage of the CC model, when the cloud compresses in direction of motion and expands perpendicular to the orbital direction.
A small fraction of the gas in the mixing zone of the cloud and atmosphere then carries angular momentum outwards, 
enabling part of the cloud to be accreted towards the center. When this outward moving gas has become very diluted 
and mixed with the 
atmosphere, the angular momentum is removed from the simulation due to our treatment to stabilize the 
unstable atmosphere.
After the pericenter passage a dilute high pressure bubble forms
(see Fig.~\ref{fig:CC01_densevol}, panel\,i), which rises buoyantly, 
thereby pushing some of the gas outward 
and the remaining gas gets less bound to the SMBH  
(Fig.~\ref{fig:bind_energy}). Model SS01 forms a disk-like structure, 
with a large amount of mass remaining in 
the high density inner region of the atmosphere.  

\begin{figure}[b]
\epsscale{1.0}
\includegraphics[width=0.95\columnwidth]{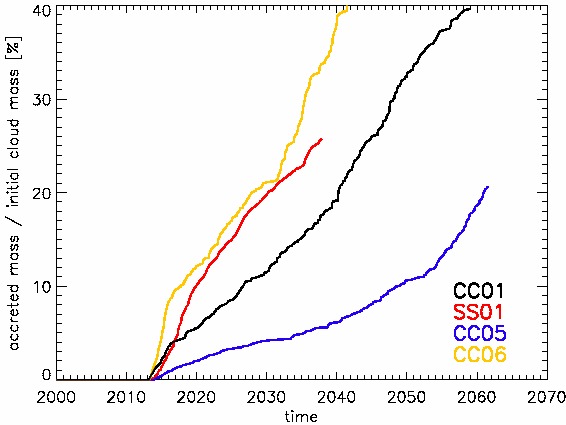}
\caption{Accreted cloud mass of the models discussed in this paper relative to their initial cloud masses. 
         See Table\,\ref{tab:simparam} for more details on the simulations.}
\label{fig:accreted_mass}
\end{figure}

\begin{figure}
\epsscale{1.0}
\includegraphics[width=0.95\columnwidth]{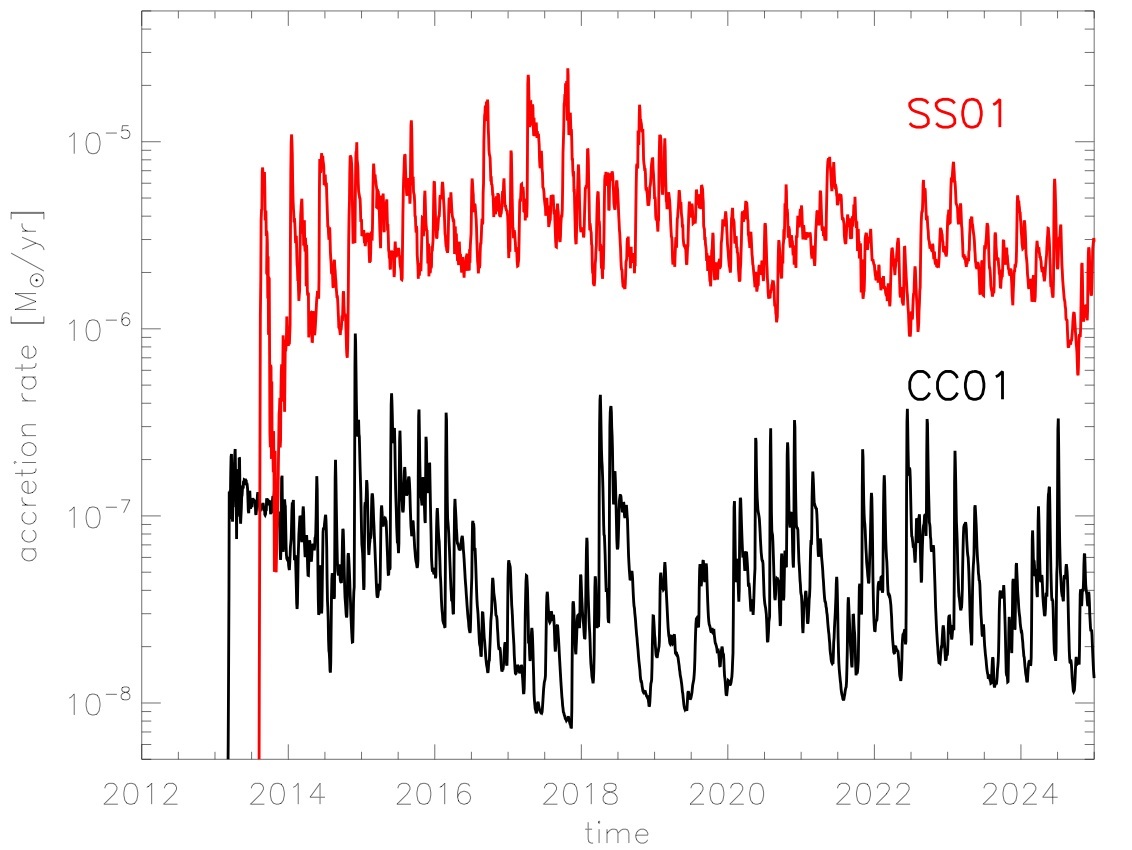}
\caption{Measured mass accretion rate through the inner boundary of our standard model CC01 (black line) and the spherical ring model SS01 (red line).}
\label{fig:accrate}
\end{figure}

During the simulations, we trace the mass inflow through the inner boundary and
the resulting total accreted mass is 
plotted against time in Fig.~\ref{fig:accreted_mass} as a fraction of the 
initial total cloud mass. Only the accreted mass originally in the cloud is taken into account.
Shortly before G2's pericenter passage in 2013.5, material moves through 
the inner boundary for the first time, but it takes roughly until 2060 to  
swallow 40\% of the cloud material in model CC01. This corresponds to an additional average accretion rate of 
$6\times 10^{-8}\,M_{\odot}\,\mathrm{yr}^{-1}$ until the year 2025. 
The time resolved accretion rate is shown in Fig.~\ref{fig:accrate} for our standard cloud and ring model. 
During the whole simulation, the time averaged accretion rate remains approximately constant.
However, hydrodynamical instabilities break up the nozzle-like stream of gas leading to clumpy accretion. 
Hence we observe deviations from the mean of up to a factor of ten with a typical time scale of the order a few months. 
Concerning the ring model SS01, the accretion through the inner boundary starts approximately half a year later
and we estimate the average accretion rate of cloud material to $4\times 10^{-6}\,M_{\odot}\,\mathrm{yr}^{-1}$ (until the 
year 2025), roughly
67 times higher compared to model CC01, again remaining roughly constant on average, but highly variable.
However, these numbers have to be interpreted with great care: Firstly, they only concern 
cloud material and secondly, in these idealized simulations, we do not take the 
back reaction of the cloud onto the ambient atmosphere into account. 
The latter could cause either enhanced accretion due to a fractional destabilization or a lower accretion rate
due to the creation of an even hotter inner bubble or feedback from the central source. 
What actually happens depends crucially on the detailed thermodynamics and the state of the
hot atmosphere in this region, which is both not very well understood and not easily accessible via
observations.

\subsection{Connection to observations and predictions for future observations}

\begin{figure}[b!]
\epsscale{1.0}
\includegraphics[width=0.95\columnwidth]{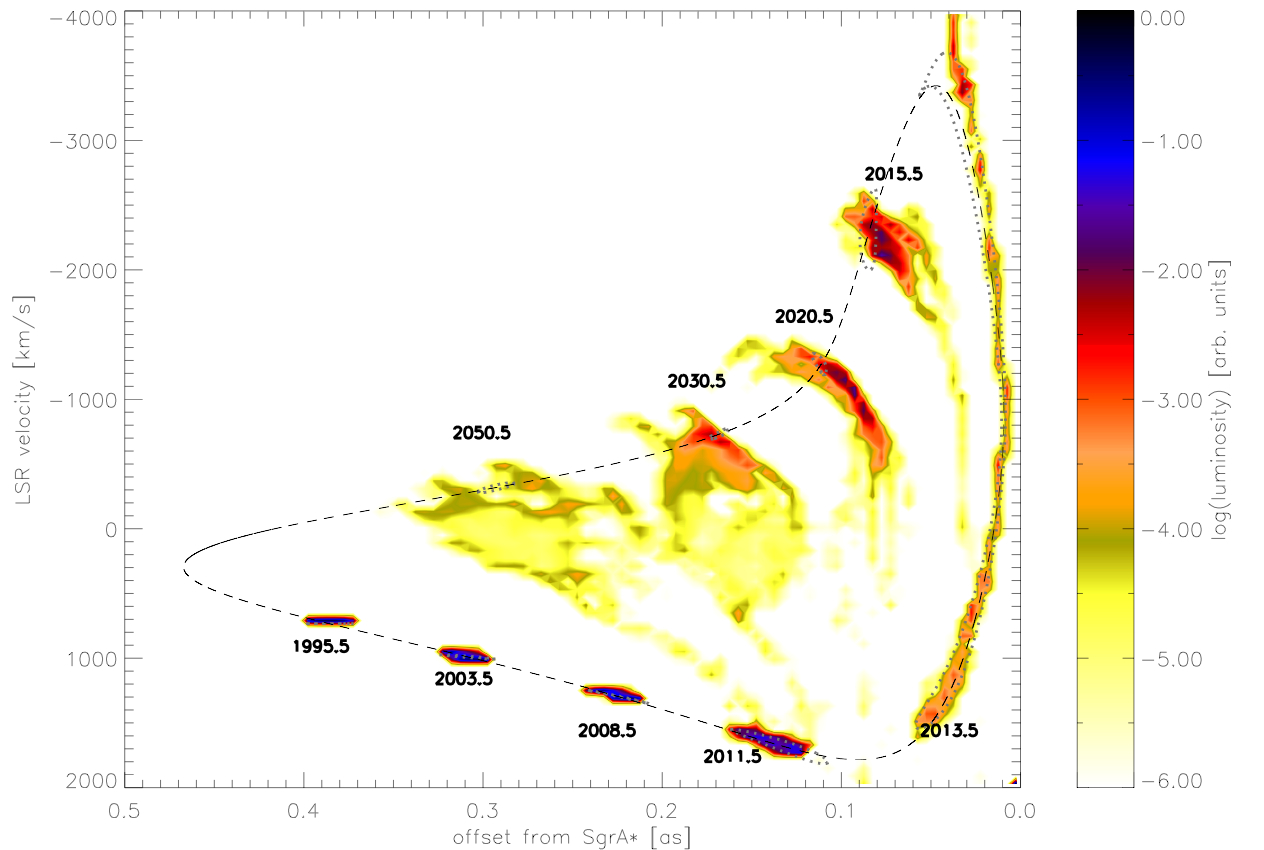}
\caption{Position-velocity diagram for the simulation CC01. The distance to Sgr\,A* -- projected on the sky -- is
plotted against the line-of-sight velocity. The colored contours display the square of a mass histogram, relative to the total
initial mass of the cloud, representative for the Br\,$\gamma$ emission. The dashed line represents 
the evolutionary path of a test particle in the center of the cloud on the orbit of G2 with the parameters 
given in Table\,\ref{tab:obsparam} for a full orbital revolution.
The dotted lines show the result of the test particle simulation.
}
\label{fig:pvtot_CC01}
\end{figure}

Fig.~\ref{fig:pvtot_CC01} displays the position-velocity (PV) diagram of the whole orbit of G2. 
It shows the projected distance of the cloud material to the black hole on sky and
the line-of-sight velocity of the cloud, 
taking the orbital angles as inferred from observations into account (Tab.~\ref{tab:obsparam}).  
Overlayed as filled contours are the data of the hydrodynamical simulation CC01 at various stages of 
its evolution, as indicated in the plot. 
It represents the square of a mass histogram of the 2D gas distribution, 
representative for the Br\,$\gamma$ emission of the cloud gas. 
The cloud starts with overall constant velocity. When moving towards the SMBH, it develops a velocity shear, 
which is the largest during pericenter passage. After being disrupted shortly after the closest approach, the 
shear decreases when the cloud gets compressed again, but never reaches the values expected from test particle
simulations, as in these stages, hydrodynamical effects dominate the evolution of the cloud.      
Nonetheless, the maximum signal is in this model always expected to arise from close to the test particle
orbit of the cloud for the whole evolutionary time probed with our simulations.

\begin{figure*}
\epsscale{1.0}
\includegraphics[width=0.9\textwidth]{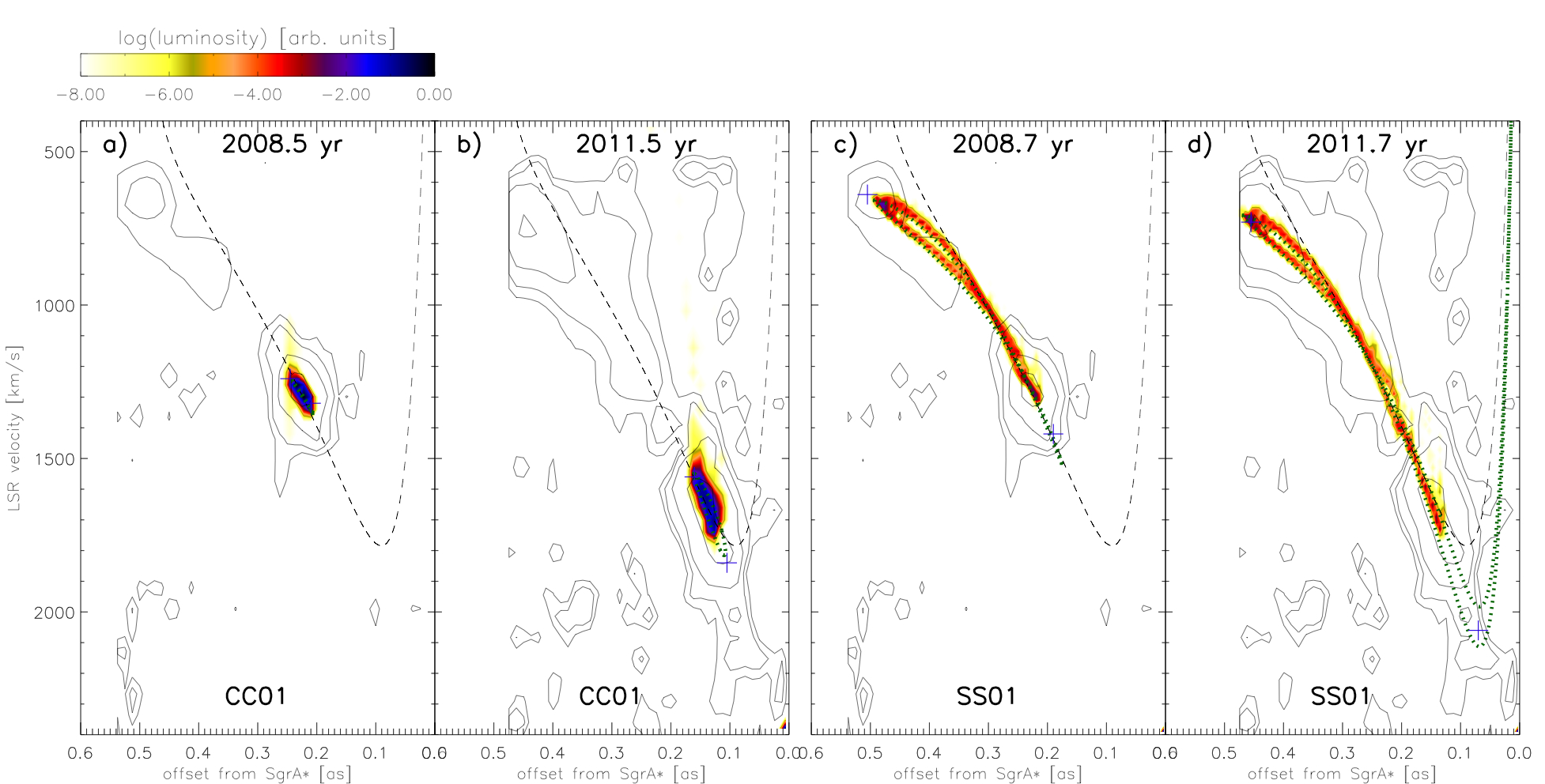}
\caption{Comparison of the position-velocity diagrams for the simulations CC01 (panels\,a,b) and SS01 (panels\,c,d) with 
observations (background contours). The distance to Sgr\,A* -- projected on the sky -- is
plotted against the line-of-sight velocity. The colored contours display the square of a mass histogram, relative to the total
initial mass of the respective cloud model. The dashed line represents 
the evolutionary path of a test particle in the center of the cloud on the orbit of G2 with the parameters 
given in Table\,\ref{tab:obsparam} for a full orbital revolution.
The green dotted lines show the result of the test particle simulation.
}
\label{fig:pvobs}
\end{figure*}

\begin{figure*}[t]
\epsscale{1.0}
\includegraphics[width=0.95\textwidth]{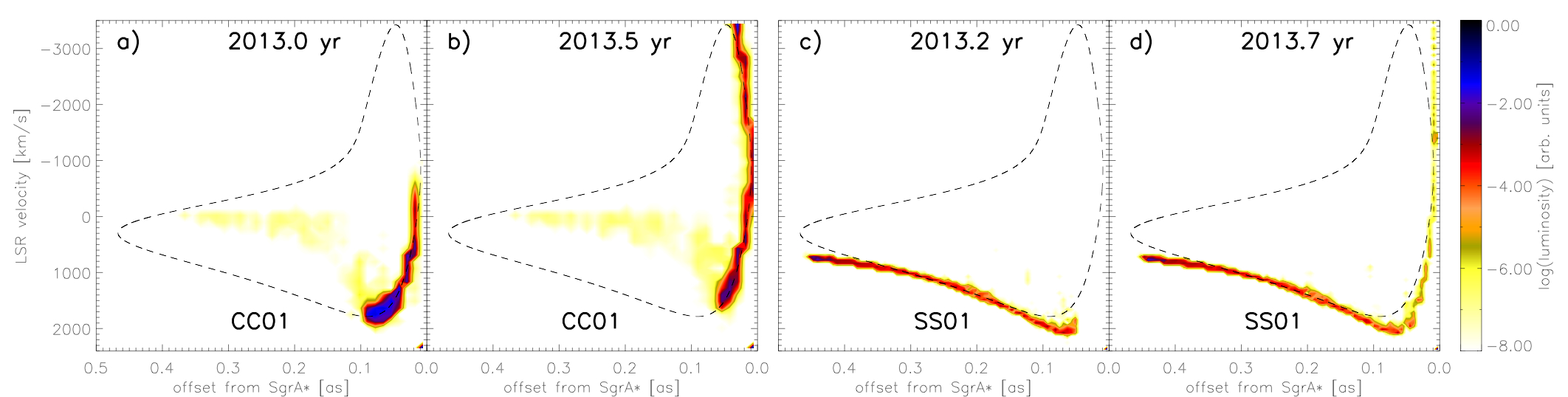}
\caption{Predictions for the distribution of the compact cloud model CC01 (panels\,a,b) and the 
spherical shell model SS01 (panels\,c,d) in the
position-velocity diagram for the year 2013. The distance to Sgr\,A* -- projected on the sky -- is
plotted against the line-of-sight velocity. The colored contours display the square of a mass histogram, relative to the total
initial mass of the respective cloud model. The dashed line represents 
the evolutionary path of a test particle in the center of the cloud on the orbit of G2 with the parameters 
given in Table\,\ref{tab:obsparam} for a full orbital revolution.
}
\label{fig:pvpredict}
\end{figure*}

We zoom into the lower part of the diagram in Fig.~\ref{fig:pvobs} for the CC model (panels\,a,b) and 
the SS model (panels\,c,d) in order to
compare to available observations and make predictions for the near-future evolution.
The compact cloud model was designed in order to give a good match
with the observed head emission in the 2008.5 and 2011.5 data, which is very well fulfilled.
In contrast to this, we interpret the observations for the case of the spherical shell model 
such that the {\it head} (G2) and {\it tail} (G2t) emission are the 
focal points of an already disrupted large cloud complex (eG2). In order to get a reasonable match with the 
observed PV diagram, this requires a spherical shell of gas, realized as a ring in our two-dimensional simulations. 
As the ring starts at its apocenter distance of $0.05\,$pc at the inner edge of the disk(s) of young stars, 
it seems obvious to interpret it as being the result of stellar evolutionary processes. 
As can be seen in Fig.\ref{fig:pvobs} (panels\,c,d),
it is in reasonable agreement with the combined observations of the {\it head} and {\it tail emission} of the cloud and 
can well explain the offset of the {\it tail} from the inferred test particle orbit of G2.
Fig.~\ref{fig:pvpredict} show our predictions for observations in 2013. 
Distinct and observable differences are visible: Whereas our compact cloud simulation (panels\,a,b) produces emission
spread around the expected location of the test particle orbit of the cloud, the spherical shell simulation 
differs significantly by extending to higher velocities up to $+2000\,\mathrm{km}\,\mathrm{s}^{-1}$ (panels\,c,d).

\begin{figure}
\epsscale{1.0}
\includegraphics[width=0.95\columnwidth]{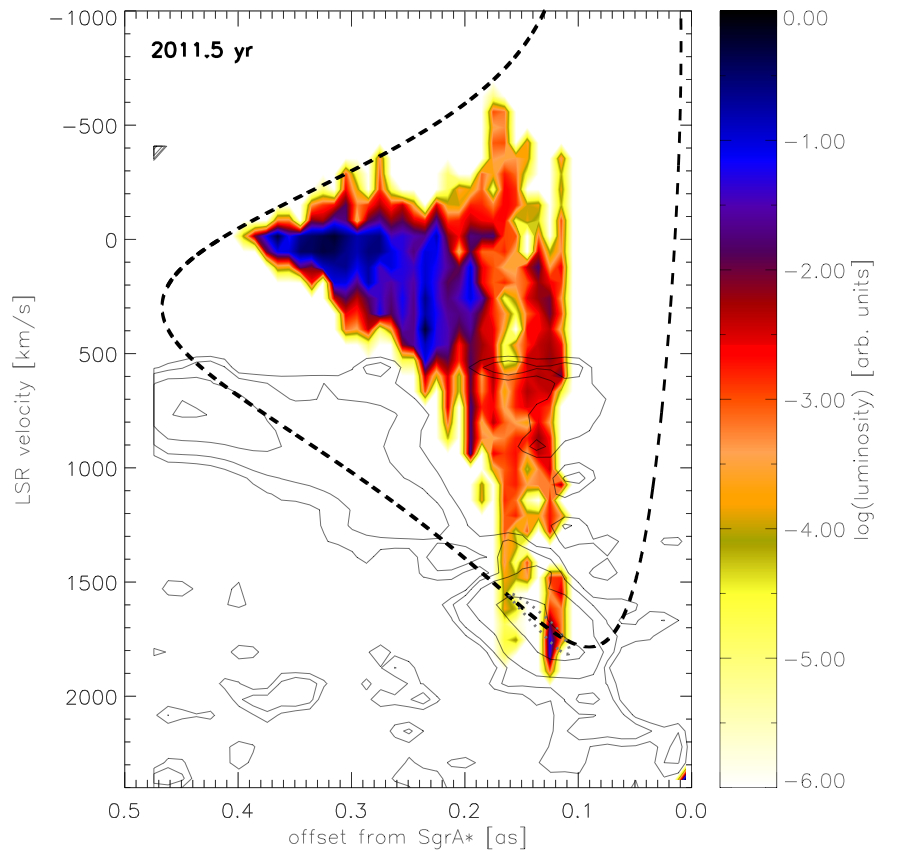}
\caption{Position-velocity diagram for the simulation CC01, where we only show the emission of the tail of the cloud,
selected by tracer values between the threshold value and 0.5. The distance to Sgr\,A* -- projected on the sky -- is
plotted against the line-of-sight velocity. The colored contours display the square of a mass histogram, relative to the total
initial mass of the cloud.
The thick dashed line represents 
the evolutionary path of a test particle in the center of the cloud on the orbit of G2 with the parameters 
given in Table\,\ref{tab:obsparam} for a full orbital revolution.}
\label{fig:pvobs_CC01_tail}
\end{figure}

In Fig.~\ref{fig:pvobs_CC01_tail} we restrict the PV-diagram to the expected emission from the hydrodynamically formed 
tail of the cloud in the {\it Compact Cloud Scenario}. 
As expected, the stripped-off gas fills the interior of the test particle curve centered on G2 (thick black dashed line). 
The earlier the gas was removed from the compact cloud, the lower its present velocity due to the ram-pressure interaction 
and mixing with  
the ambient hot medium. The turbulent motions of this low-density material can be seen in the large width of 
the distribution in velocity direction, which even extends to velocities with the opposite sign.  
The contours vertically offset from the actual location of G2 could be related to the stripped-off gas from the front part of the cloud.
Fig.~\ref{fig:pvobs_CC01_tail} shows that what was often referred to as the {\it tail} of G2 in the 
observed PV diagram, cannot be explained by stripped material from the compact cloud, causing problems for the Compact Cloud model
or any model where the cloud is the result of a compact invisible source
of gas ({\it Compact Source Scenario}, 
see Section\,\ref{sec:intro} and \citealp{Murray_Clay_11}).
Instead, the so far undiscussed contours at low distances to Sgr~A* and low velocities could be related to a 
hydrodynamically stripped tail of gas.
In contrast, our {\it Spherical Shell Scenario} is qualitatively able to explain the  whole structure of the observed PV-diagram.

A second promising possibility to distinguish between different model predictions is via X-ray and IR-observations. 
In Sect.~\ref{sec:accretion}, we discussed the mass accretion rate through the inner boundary of our domain.
\citet{Baganoff_03} observationally determine an accretion rate of $10^{-6}\,M_{\odot}\,\mathrm{yr}^{-1}$ at the Bondi radius, 
located at $10^5$ Schwarzschild radii \citep{Genzel_10}. We use the scaling relation

\begin{eqnarray}
  \dot{M}_{\mathrm{in}} = \dot{M}_{\mathrm{Bondi}}\,\left( \frac{r_{\mathrm{in}}}{r_{\mathrm{Bondi}}} \right)^s,
\end{eqnarray}

where $r_{\mathrm{in, Bondi}}$ is the inner radius or the Bondi radius
and $\dot{M}_{\mathrm{in, Bondi}}$ is the mass accretion rate at the inner boundary of our domain or the Bondi radius, respectively
\citep[e.\,g.\,][]{Blandford_99,Igumenshchev_03} and $s=0.27$ \citep{Yuan_03} to calculate the accretion rate expected
at the inner boundary of our domain, which is located at roughly 1500 Schwarzschild radii distance from the SMBH. It 
results in a value of $3\times 10^{-7}\,M_{\odot}\,\mathrm{yr}^{-1}$. 
As discussed in Sect.\,\ref{sec:accretion}, the matter streams through our inner boundary in form of small droplets or
filaments. Given the proximity of the SMBH, the evaporation timescale is very short (see equation \ref{equ:evaptime}).
Hence our first assumption is that the cloud material will evaporate on its further way in and join the hot accretion flow.
To estimate the effect on the observable luminosity of Sgr\,A*, we use the spectral energy distributions derived for 
the ADAF models of \citet{Yuan_04}. We interpolate their SEDs for various mass accretion rates shown in their Fig.\,5 
to match the average mass accretion rates of the flow through the inner radii of our models and compute the 
IR ($0.8-2.5\micron$) and X-ray (2-8\,keV) luminosity. If the mass accretion proceeds in this mode, we do not expect 
a significant boost of the X-ray or IR luminosity for the CC01 model, as even the spikes ten times above the average
accretion rate would only lead to a factor of 1-2 increase of the IR and X-ray signal.
In contrast to this, model SS01 would result in an X-ray luminosity of $4\times 10^{34}\,\mathrm{erg}\,\mathrm{s}^{-1}$, corresponding to 
a boost of a factor of roughly 80 compared to the quiescent X-ray luminosity of 
Sgr\,A* ($4\times 10^{32}\,\mathrm{erg}\,\mathrm{s}^{-1}$ derived from 
\citealp{Yuan_04}, Fig.~\,5). The expected IR luminosity amounts to $5\times 10^{35}\,\mathrm{erg}\,\mathrm{s}^{-1}$, roughly the same factor
above the assumed quiescent value ($7\times 10^{33}\,\mathrm{erg}\,\mathrm{s}^{-1}$). Concerning the spikes of the accretion rate distribution
(roughly a factor of ten above the average accretion rate), we calculate boost factors of roughly $3\times 10^4$ in 
the X-ray and $6\times 10^3$ in the IR. These high boost factors would enable us to test the validity of our spherical 
shell scenario in the coming years. 
However, another possibility (we are unable to test with our current models) is that the cold, disrupted cloud significantly 
affects the thermodynamic structure of the hot accretion flow close to the GC. 
This might then prevent the complete thermal evaporation of the cloud fragments and enable
the formation of a thin and relatively cold accretion disk. 
To get a first rough estimate of the expected signal, we use the average mass transfer rate through the 
inner boundary of our domain (Fig.~\ref{fig:accrate}) and assume it is representative for the 
steady-state mass accretion rate through such an inner accretion disk, which we assume forms instantly. 
Given this assumption, the bolometric luminosity of the disk can be estimated to be \citep{Krolik_99}:

\begin{eqnarray}
  L_{\mathrm{bol}}=\frac{1}{12}\,\dot{M}\,c^2
\end{eqnarray} 

This results in an average bolometric luminosity of $8\times 10^4\,L_{\odot}$ and $5\times 10^6\,L_{\odot}$ for the cloud model 
and the ring model respectively (averaged over a time period from the start of the accretion until the year 2025). 
In order to derive the observable X-ray luminosity, we 
use the accretion disk spectral energy distribution constructed in \citet{Schartmann_05} from a combination of observations and 
simulations. This results in averaged values of $L_{\mathrm{2-8keV}}=4\times10^{33}\mathrm{erg}\,\mathrm{s}^{-1}$ 
and $L_{\mathrm{2-8keV}}=3\times10^{35}\mathrm{erg}\,\mathrm{s}^{-1}$ for the 
cloud and ring model in the energy range between two and eight keV and $3\times 10^{37}\mathrm{erg}\,\mathrm{s}^{-1}$ 
and $10^{39}\,\mathrm{erg}\,\mathrm{s}^{-1}$ for 
the two models in the IR, the latter being orders of magnitude above the current quiescent luminosity.
Which fraction of the cloud material will accrete in this mode is currently unclear and the detailed prediction of the resulting SED 
is clearly beyond the scope of this publication, but might influence these values significantly.
However, a similar increase of the luminosity compared to the current state 
might in turn influence the RIAF solution itself \citep{Yuan_04} as well as the 
dynamical evolution of the cloud via its radiation pressure 
interaction, as has for example been investigated by \citet{Schartmann_11}.
More detailed modelling of the physical processes at work as well as the 
hot atmosphere is necessary to 
determine the expected light curve, triggered by the future evolution of the cloud.
Thermal conduction might be one of the key players here as the cloud evaporation timescale
becomes very short in this phase.

\subsection{Influence of the equation of state}
\label{sec:eos}

\begin{figure}[b]
\epsscale{1.0}
\includegraphics[width=0.95\columnwidth]{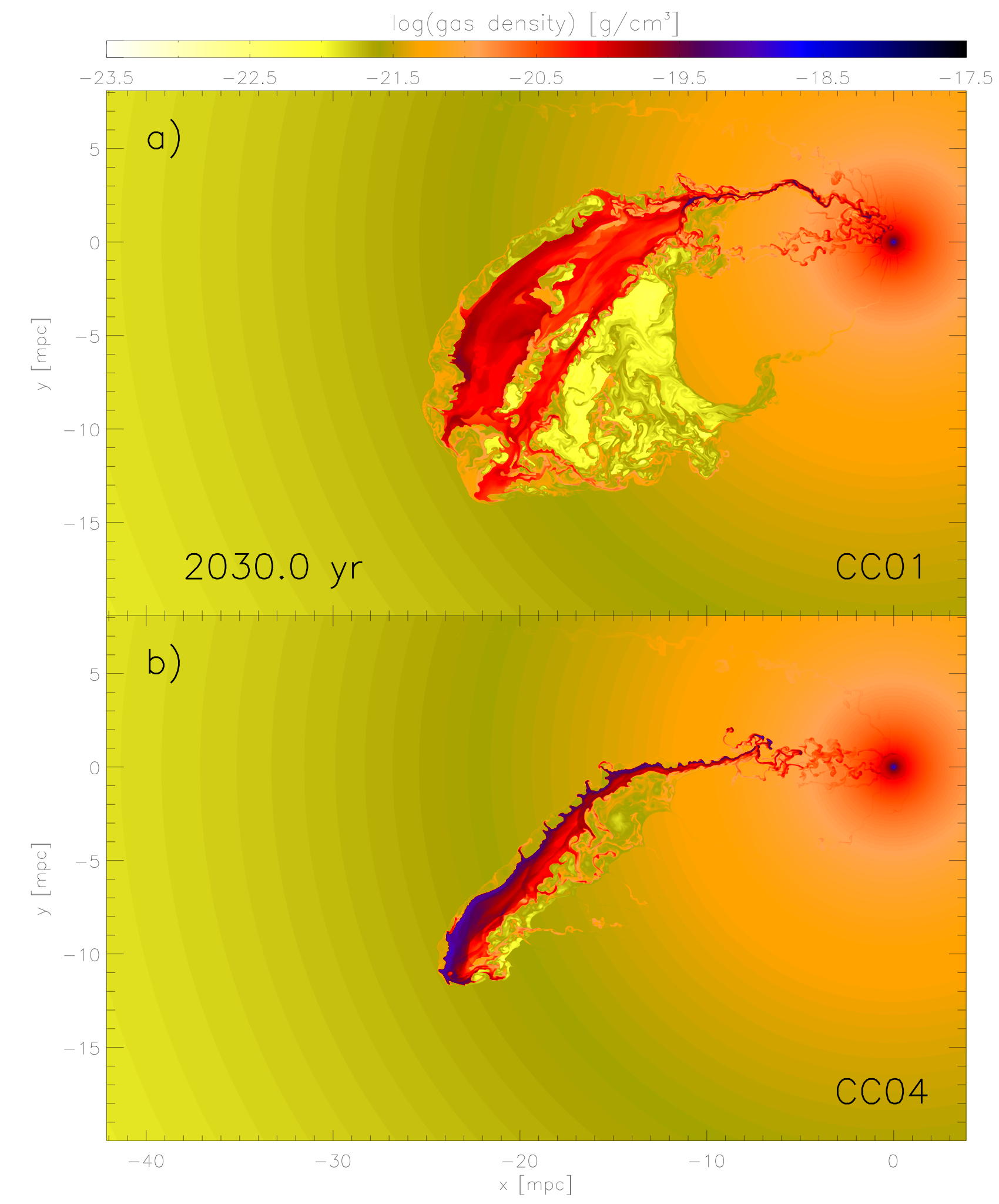}
\caption{Comparison of the density of the cloud in 2030 for our standard model (panel\,a, CC01) and the 
adiabatic test simulation (panel\,b, CC04).}
\label{fig:CC01_adcomp}
\end{figure}

\begin{figure*}
\epsscale{0.75}
\includegraphics[width=0.95\textwidth]{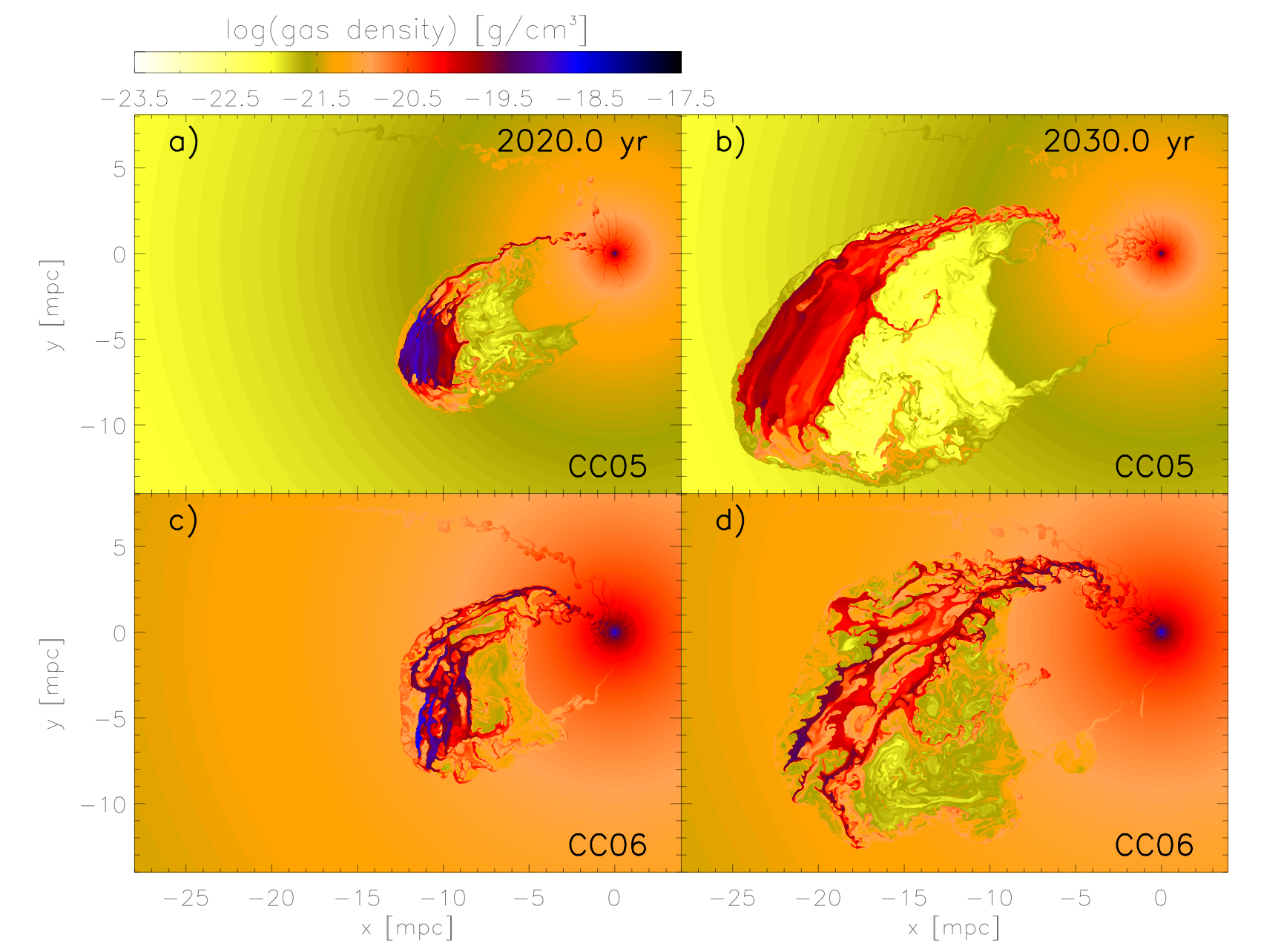}
\caption{Comparison of two compact cloud models with a factor of two decreased density of the atmosphere
         (panel\,a,b, CC05) and a factor of two increased density of the atmosphere (panel\,c,d, CC06) with respect to
         our standard model.}
\label{fig:atdens_study}
\end{figure*}

A test simulation has been run with an adiabatic equation of state (simulation CC04). This represents
inefficient external cooling and heating sources, but takes adiabatic cooling and heating via expansion and 
contraction into account. The early evolution of the density distribution is very similar in both 
cases. Major differences only arise after the pericenter passage. A comparison of the state in the 
year 2030 is presented in Fig.~\ref{fig:CC01_adcomp}.
We find that in simulation CC01, the pressure inside the remaining cloud gas 
is larger compared to the adiabatic case CC04. 
The reason is that the gas in the adiabatic simulation is able to cool when the cloud expands and gets
disrupted after the pericenter passage. 
This effect is also visible in the formation of a low density but high pressure
region in downstream direction of the cloud, which is less pronounced in the adiabatic simulation.
The lower temperature and pressure in the adiabatic case lead to 
a more filamentary cloud with a stronger density enhancement, whereas in the other case, the larger 
pressure support leads to a more uniform density distribution, spreading over a larger area.
This example shows again that the early evolution of the cloud can easily be understood with simple hydrodynamical
simulations and we can trust the predictions for the near-future observations.
However, the late-time evolution -- after tidal disruption has severely affected the cloud structure --
significantly depends on the detailed physical processes and the structure of the surrounding atmosphere. 
For an in-depth understanding of the long-term fate of the cloud, more sophisticated (magneto)-hydrodynamical simulations
are required.

\subsection{Influence of the atmospheric density distribution}

Our knowledge of the temperature and density distribution of the hot 
atmosphere in the Galactic Center region 
is based on theoretical models, which are consistent with 
the small number 
of data points observed with the Chandra satellite and the rotation measures (see above). 
Hence, it is interesting to ask whether the current and future evolution
of the cloud will enable us to probe the hot atmosphere.
A first impression of the effect of the density distribution of the ambient 
atmosphere on the cloud evolution was given in Sect.~\ref{sec:testpart_ss}.
As was shown there under the assumption of initial pressure equilibrium, 
the starting date would be changed by five to ten years if the
atmospheric density distribution would be changed 
by a factor $f_{\mathrm{hot}}$ of 0.5 (CC05) or 2.0 (CC06).
However, the absolute values of the preferred origin date of the cloud depend
on the not very well known total mass of the cloud and
hence our results should be taken with care. 
Here, we briefly discuss the hydrodynamical realizations of these two
best-fit models from the test particle simulations for the 
{\it Compact Cloud Scenario}. 
As expected, both clouds are in good agreement with the observed PV diagrams in 
2008.5 and 2011.5, but the late-time evolution of the density distribution 
changes significantly, as can be seen in Fig.~\ref{fig:atdens_study}. 
The growth of hydrodynamical instabilities scales with the density 
contrast of the cloud/atmosphere system and the importance of ram pressure
effects increases with the density of the environment (see Sect.~\ref{sec:involv_phys}).
The earlier onset of KHI in the high-density case 
-- already during the pericenter passage -- leads to a more filamentary
cloud boundary. The interior is affected as well due to the partly disruption 
and the later buckling of the cloud.
Gas transfer through the inner boundary starts roughly at the same time in both cases. 
However, the reduced ram pressure for the case of simulation CC05 leads to a smaller accretion rate
(blue graph in Fig.~\ref{fig:accreted_mass}) compared to the high density atmosphere (CC06), which displays a steeper increase of the 
accretion rate with time (yellow curve in Fig.~\ref{fig:accreted_mass}).  
Concerning the evolution in the PV-diagram, model CC06 spreads over a larger region during the late evolution due 
to the faster disruption of the cloud in the high density case.

\section{Discussion}
\label{sec:discussion}

Numerically, the most critical point of the simulations is the treatment of the atmosphere.
As already discussed in Sect.~\ref{sec:mod_atmosphere}, the ADAF solution which fits 
the Chandra observations of the hot atmosphere best is  
unstable to convection and cannot be modeled in a simple way. 
In order to enable a first investigation of the origin and further evolution of the 
cloud, we artificially stabilize the atmosphere and make several simplifying assumptions: 
(i) We neglect the accretion flow of the diffuse atmosphere 
towards the center and set up the central hot gas distribution
in hydrostatic equilibrium. (ii)  We suppress the growth of disturbances in the atmosphere by 
re-setting those cells within the atmosphere, which have not 
interacted with the cloud yet. This is necessary in particular close to the central boundary condition, where disturbances 
due to the Cartesian grid start to increase due to the entropy gradient.
To avoid these complications, we use a passive tracer field. 
Cloud material is lost in those regions of the computational grid where the 
fraction of cloud gas relative to atmospheric gas is below a threshold of
$10^{-4}$.
This leads to a roughly linear increase of the mass loss starting around the time of the pericenter passage in
the standard model CC01. At the end of the simulation in the year 2060, roughly 1.3\% of the initial cloud mass 
is removed from the simulation. 
This also means that the angular momentum and kinetic energy, which the cloud transfers to the atmosphere is 
finally taken out of the simulation. A critical discussion of the influence of the threshold value for the 
tracer is given in Sect.~\ref{sec:tracer_eff}.
Further limitations of our simulations are the negligence of thermal conduction, which will be important 
at least in the late phase of the evolution of the cloud (Sect.~\ref{sec:therm_cond}) and
the restriction of the simulations to two dimensions.
Perpendicular to the modelled orbital plane, we expect the cloud to collapse under the 
influence of the gravitational force of the black hole, forming a thin disk and later a thin filament close
to pericenter. After passing the black hole, the cloud is expected to expand again in vertical direction due
to the overpressure with respect to the surrounding medium. Given the larger contact surface with the hot ambient medium 
in rest, enhanced stripping due to KHI is expected, which is most important for the late time shredding of 
the cloud after pericenter passage. In contrast to this, we expect less ram pressure interaction in the 3D case,
which might affect the angular momentum exchange with the atmosphere and hence slightly alter the mass transfer towards the center.

\newpage

\subsection{Resolution effects}
\label{sec:res_eff}

\begin{figure}
\epsscale{1.0}
\includegraphics[width=0.95\columnwidth]{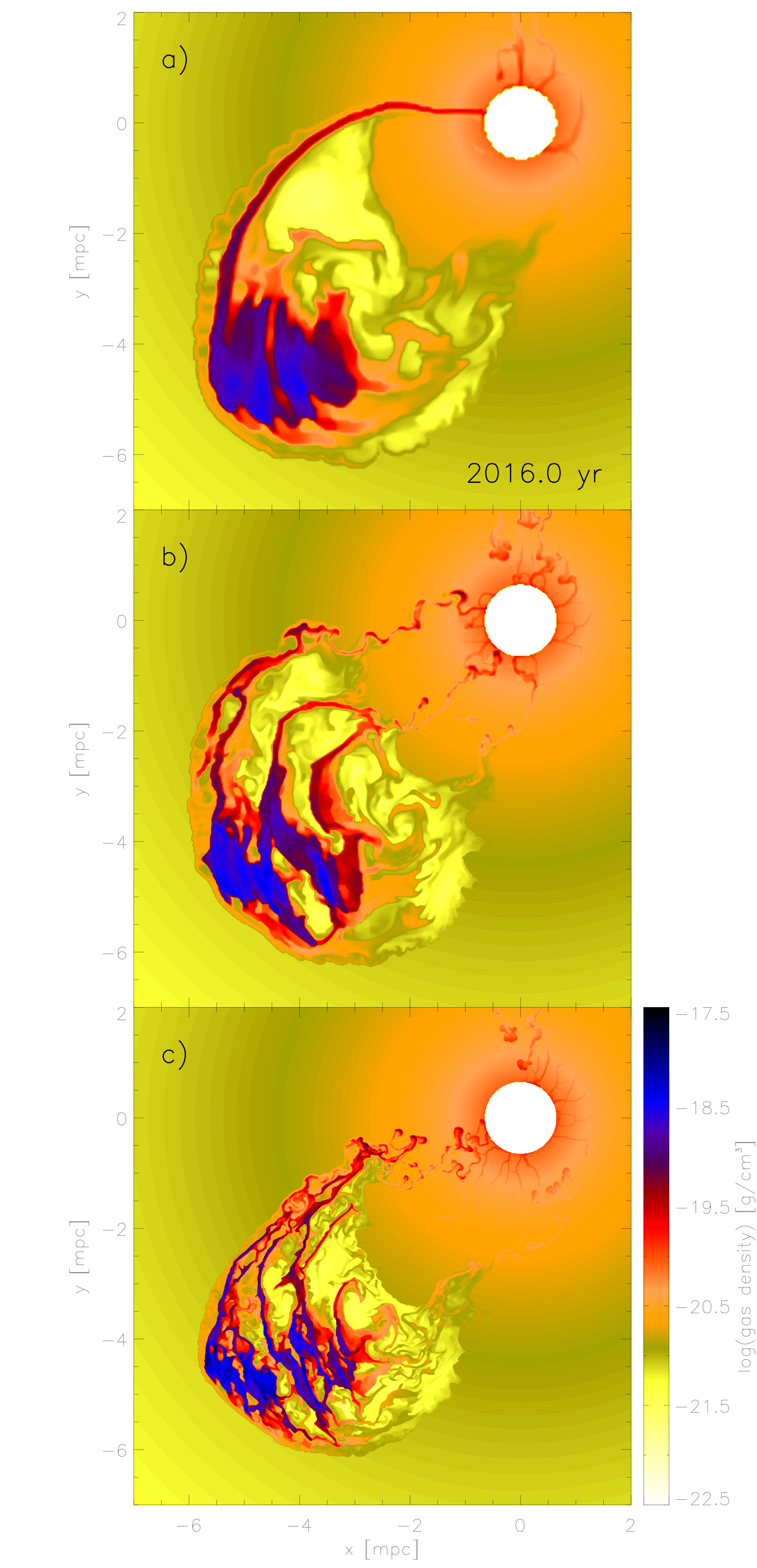}
\caption{Resolution study of our standard model. Displayed is our standard resolution (panel\,b, CC01), half the 
resolution (panel\,a, CC07) and twice the resolution (panel\,c, CC08).}
\label{fig:resolution_study}
\end{figure}

The effects of resolution on the density evolution of our simulations are displayed in 
Fig.~\ref{fig:resolution_study}. The snapshots are taken from the compact cloud simulation starting
on G2's orbit in the year 1995.5. Panel\,b displays our standard resolution, whereas we 
decreased the resolution in both axis-directions by a factor of two in panel\,a and increased 
the resolution by a factor of two for the simulation displayed in panel\,c. The snapshots 
correspond to the year 2016, shortly after the pericenter passage. 
The basic dynamical evolution as well as the disruption due to tidal forces
is very comparable. However, the expected resolution effects are visible in the detailed evolution
of the hydrodynamical instabilities. This leads to a stronger disruption of the nozzle-shaped inflow
of gas towards Sgr\,A* with increased resolution. 
In the highest resolution simulation, the stream has broken up already and
the mass-flux through the inner boundary is clumpy, compared to the smooth accretion of gas through
the nozzle in our low resolution run. Our standard resolution run already shows some clumpy accretion, 
which also increases at later stages of the evolution (Fig.~\ref{fig:CC01_densevol}, lower row).
The mentioned accretion rates through the inner boundary differ by about 10\% between the highest and
the lowest resolution in our study.

\subsection{Influence of the choice of the tracer threshold value}
\label{sec:tracer_eff}

\begin{figure}
\epsscale{1.0}
\includegraphics[width=0.95\columnwidth]{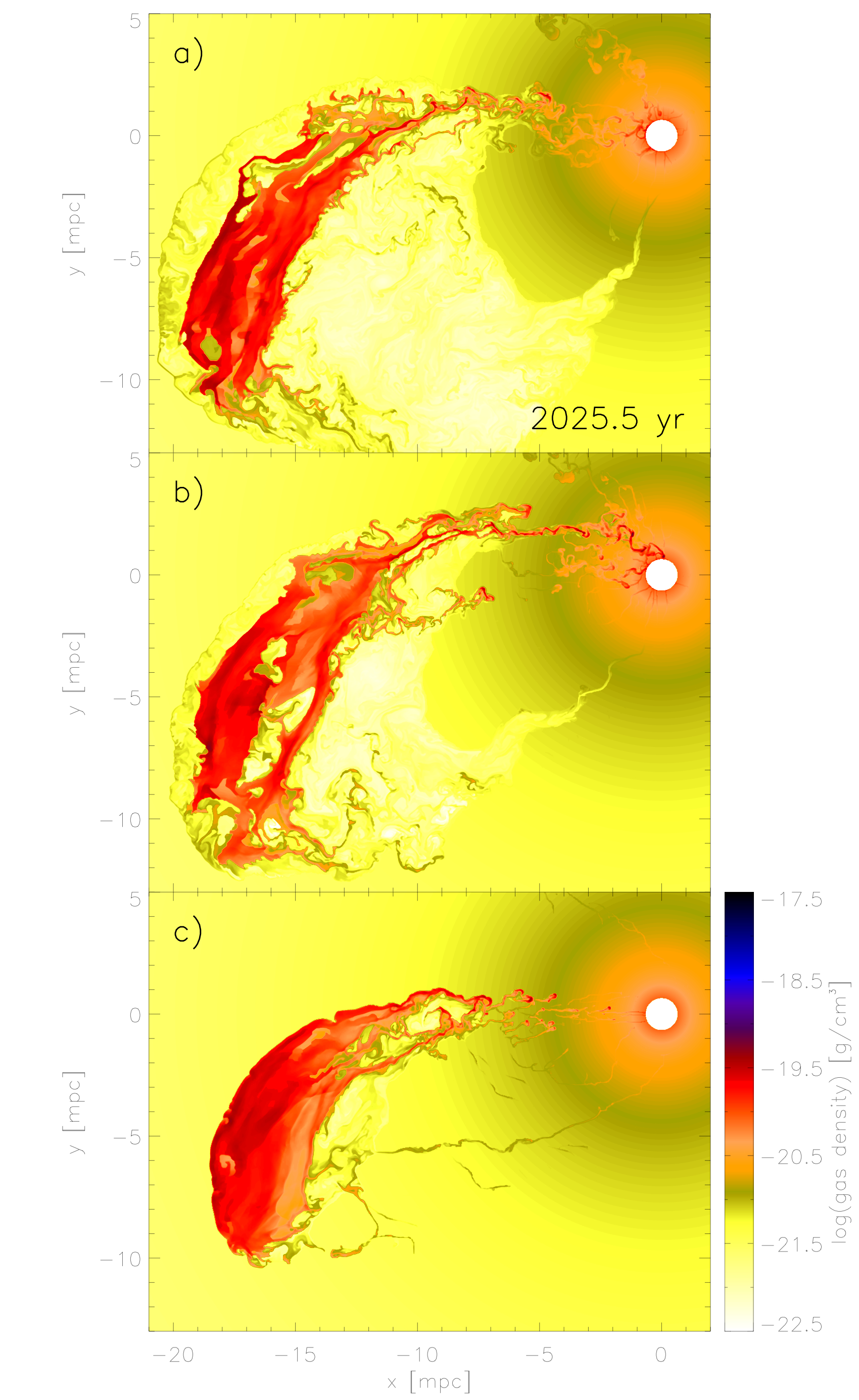}
\caption{Studying the influence of the threshold value of the tracer field for artificially stabilizing 
the atmosphere. Displayed is a value of $10^{-6}$ (panel\,a),
$10^{-4}$ (panel\,b, our standard model) and $10^{-2}$ (panel\,c).}
\label{fig:tracer_study}
\end{figure}

Fig.~\ref{fig:tracer_study} displays the influence of the tracer threshold value. This is done for our
simulation starting on G2's orbit in the year 1995.5 (model CC01) and we analyse the state of the simulation in the 
year 2025.5, as only the late-time evolution shows significant differences. 
From panel\,a to c, the tracer threshold value increases from $10^{-6}$ (panel\,a)
to $10^{-4}$ (panel\,b) and $10^{-2}$ (panel\,c).  
Again, as in the resolution study, the basic dynamical evolution is unaffected by the choice of the threshold 
value. 
However, in this series, two systematic effects are visible:
(i) For too high threshold values (Fig.~\ref{fig:tracer_study}c), 
    the shape of the outer boundary in upstream direction changes slightly. As soon as the cloud mixes with 
    the surrounding medium, gas is removed from the simulation. This artificially suppresses instabilities.
(ii) For too low threshold values, the disturbed atmosphere in downstream direction of the cloud gets 
     susceptible to the growth of convectively unstable regions. Bubbles are formed behind the cloud as 
     well as in the disturbed atmosphere along the traversed orbit of the cloud. 
     Quantifying the effects of the de-stabilization of a realistic atmosphere and its back-reaction onto 
     the further evolution of the cloud is beyond the scope of these idealized simulations. 
Given the considerations briefly discussed here, 
we decided to use a tracer threshold value of $10^{-4}$.

\subsection{Possible influence of magnetic fields on the evolution of the cloud}
\label{sec:magfields}

Strong magnetic fields of the order of $100\,\mu$G (with a lower limit of $50\,\mu$G on 400\,pc scales) have 
been observationally inferred near the Galactic Center  
\citep{Yusef_Zadeh_87,Morris_89,Crocker_10}. Field strengths of this magnitude might significantly influence 
the dynamics and physical interactions occuring in the interaction region of 
the cloud. Several effects might change the dynamics and appearance of the cloud: 
(i) magnetic pressure inside the cloud and of the hot atmosphere changes the total pressure balance and
hence the size along the cloud's orbit, (ii) magnetic tension forces can lead to a confinement of the cloud 
\citep[e.~g.~][]{Krause_12}, thereby alleviating the problem of the too fast tidal disruption, 
(iii) depending on field morphology, the Kelvin-Helmholtz-Instability can be significantly suppressed, again
leading to a longer survival of the cloud and significant changes in the late time evolution and 
(iv) magnetic fields lead to anisotropic thermal conduction, altering the cloud evaporation time scale.
Hence it will be necessary to consider magnetic fields in future simulations determining the fate 
of the observed cloud. The problem is that this adds additional degrees of freedom as neither the 
strength of the fields nor its morphology is currently known in the sub-parsec surrounding of Sgr\,A*.

\section{Conclusions}
\label{sec:conclusions}

In this paper, we present a set of idealized hydrodynamical simulations to study the origin and the fate
of the recently discovered cloud complex in the Galactic Center. 
Building up on analytical estimates in paper\,I, we concentrate on two possible scenarios: 
(i) the {\bf Compact Cloud Scenario}, where the cloud is assumed
to have formed in the recent past on the orbital path of the G2 component of the cloud and 
(ii) the {\bf Spherical Shell Scenario}, in which the observations 
are interpreted as an already disrupted spherical shell (or ring, respectively in two dimensions). 
The early evolution until close to the pericenter is dominated by tidal interaction in both cases. 
Shortly after pericenter passage, when the cloud is already tidally disrupted, secondary hydrodynamical instabilities 
as well as ram pressure become important, leading to some angular momentum redistribution and accretion towards 
the center.
Nonetheless, the cloud keeps a large fraction of its initial angular momentum, which efficiently 
hinders direct accretion onto Sgr\,A*. Accretion is expected to occur via a thin nozzle-like feature connecting
the cloud with the SMBH, as a result of continuous ram-pressure interaction of the upstream boundary of the cloud
with the dense inner part of the atmosphere. 
Hence -- depending on the destabilization of this stream (e.~g.~by thermal conduction) --
there might be enhanced cold gas accretion within the next years, which might boost the 
luminosity of Sgr~A* in an observable way. However, according to our simulations,  
a large fraction of the cloud will remain on a similar orbit compared to the test particle orbit and 
lead to a steady inflow of gas towards the Galactic Center.
In contrast to the early evolution of the cloud, the post-pericenter stage depends sensitively on the detailed modelling of physical 
processes as well as the atmospheric density distribution, necessitating more detailed hydrodynamical simulations 
to predict the (far) future evolution of the cloud complex. 
A more realistic treatment of the unstable nature of the atmosphere might influence our derived accretion rates substantially, when 
taking the back-reaction of the cloud into account. 

In summary, we currently favor the {\bf Spherical Shell Scenario} for the following reasons:
(i)   It allows the cloud to be started within the range of the disk(s) of young stars. 
(ii)  Being a spherical shell of gas further strengthens the notion of it being the result of 
      stellar evolutionary processes. 
(iii) The model is able to explain the major part of the structure (eG2) seen in the observed position-velocity diagram,
      whereas the {\it Compact Cloud Scenario} only captures the dominant part of it (G2).
(iv)  Tidally stripped gas cannot account for the G2t component of the observed PV-diagram, which disfavors 
      the {\it Compact Cloud Scenario} and any 
      {\it Compact Source Scenario}, but a so far undiscussed component of the PV-diagram could be directly related to it.


\acknowledgments

We are grateful to Romain Teyssier for helpful discussions 
and an anonymous referee for helpful comments. 
This work was supported by the Deutsche Forschungsgemeinschaft priority program 1573
("Physics of the Insterstellar Medium").


\bibliographystyle{apj}
\bibliography{apj-jour,literature}


\end{document}